\documentclass[reprint,superscriptaddress,nofootinbib,amsmath,amssymb,aps,pra,floatfix]{revtex4-1}

\usepackage{amssymb}
\usepackage{graphicx}

\usepackage{amsmath,graphicx,soul,subcaption,balance}
\usepackage{balance}
\usepackage{parskip}

\usepackage{mathptmx}

\usepackage{multirow} 

\usepackage{color}
\usepackage{soul,mathrsfs}
\usepackage[utf8]{inputenc}
\usepackage[T1]{fontenc}
\usepackage{array}

\usepackage[noend]{algpseudocode} 
\usepackage{algorithm} 
\usepackage{float}
\usepackage{relsize} 
\usepackage{tikz} 
\usepackage{DejaVuSans} 
\DeclareMathAlphabet{\mathcal}{OMS}{cmsy}{m}{n} 
\usepackage{mathtools} 
\usepackage{bbm} 
\usepackage{booktabs} 
\usepackage{aligned-overset} 
\usepackage{enumitem} 
\usepackage{hyperref}
\usepackage[hyphenbreaks]{breakurl}
\usepackage{bm} 

\usepackage{graphicx}
\usepackage{tikz}
\usetikzlibrary{positioning}
\usepackage{tikzscale}
\usepackage{import}
\usetikzlibrary{quotes,arrows.meta}
\usetikzlibrary{positioning}

\usepackage{Box}
\usetikzlibrary{backgrounds,calc,shadings,shapes.arrows,arrows,shapes.symbols,shadows,positioning,decorations.markings,backgrounds,arrows.meta}
\makeatletter
\newcommand*\bigcdot{\mathpalette\bigcdot@{.5}}
\newcommand*\bigcdot@[2]{\mathbin{\vcenter{\hbox{\scalebox{#2}{$\m@th#1\bullet$}}}}}
\makeatother

\definecolor{darkblue}{HTML}{1f4e79}
\definecolor{lightblue}{HTML}{00b0f0}
\definecolor{salmon}{HTML}{ff9c6b}
\definecolor{dodgerblue}{rgb}{0.12, 0.56, 1.0}
\definecolor{frenchblue}{rgb}{0.0, 0.45, 0.73}
\definecolor{green(pigment)}{rgb}{0.0, 0.65, 0.31}
\definecolor{macaroniandcheese}{rgb}{1.0, 0.74, 0.53}
\definecolor{arylideyellow}{rgb}{0.91, 0.84, 0.42}
\definecolor{pansypurple}{rgb}{0.47, 0.09, 0.29}
\definecolor{glaucous}{rgb}{0.38, 0.51, 0.71}
\definecolor{hanblue}{rgb}{0.27, 0.42, 0.81}
\definecolor{newblue}{rgb}{0.56, 0.67, 0.85}
\definecolor{newgreen}{rgb}{0.67, 0.82, 0.57}
\definecolor{fireenginered}{rgb}{0.81, 0.09, 0.13}
\definecolor{lightwheat}{rgb}{0.98, 0.93, 0.85}
\definecolor{navyblue_mod}{rgb}{0.0, 0.23, 0.72}

\begin{document}

\title{Understanding Concepts in Graph Signal Processing for Neurophysiological Signal Analysis}

\author{Stephan Goerttler}
  \email{Correspondence to: goerttlers@uni.coventry.ac.uk}
  \affiliation{Centre for Computational Science and Mathematical Modelling, Coventry University, Coventry CV1 2JH, UK 
  }

\author{Fei He}
  \affiliation{Centre for Computational Science and Mathematical Modelling, Coventry University, Coventry CV1 2JH, UK
}

\author{Min Wu}
  \affiliation{Institute for Infocomm Research, Agency for Science, Technology and Research (A*STAR), 138632, Singapore
}

\begin{abstract}
Multivariate signals, which are measured simultaneously over time and acquired by sensor networks, are becoming increasingly common. The emerging field of graph signal processing (GSP) promises to analyse spectral characteristics of these multivariate signals, while at the same time taking the spatial structure between the time signals into account. A central idea in GSP is the graph Fourier transform, which projects a multivariate signal onto frequency-ordered graph Fourier modes, and can therefore be regarded as a spatial analog of the temporal Fourier transform. This chapter derives and discusses key concepts in GSP, with a specific focus on how the various concepts relate to one another. The experimental section focuses on the role of graph frequency in data classification, with applications to neuroimaging. To address the limited sample size of neurophysiological datasets, we introduce a minimalist simulation framework that can generate arbitrary amounts of data. Using this artificial data, we find that lower graph frequency signals are less suitable for classifying neurophysiological data as compared to higher graph frequency signals. Finally, we introduce a baseline testing framework for GSP. Employing this framework, our results suggest that GSP applications may attenuate spectral characteristics in the signals, highlighting current limitations of GSP for neuroimaging.

\end{abstract}

\maketitle

\section*{Acknowledgment}
\begin{quote}
This is a preprint of the following chapter: Stephan Goerttler, Fei He, and Min Wu, Understanding Concepts in Graph Signal Processing for Neurophysiological Signal Analysis, published in \textit{Machine Learning Applications
in Medicine and Biology}, edited by Ammar Ahmed and Joseph Picone, 2024, Springer, reproduced with permission of Springer Nature Switzerland AG. The final authenticated version is available online at: \url{https://dx.doi.org/10.1007/978-3-031-51893-5_1}.
\end{quote}

\section{Introduction}
\label{sec:intro}
Multivariate signals consist of multiple time signals that are acquired simultaneously using an array of spatially separated sensors. The lowering cost of sensors has made multivariate signals ubiquitous, and tools to process these signals are urgently needed.
One instructive example of a multivariate signal from the field of geophysics is a temperature measurement at different geographic locations. Here, each location measures a time signal, which gives information about the temperature fluctuations at this particular location over time. We can also look at the temperature measurements across all locations at any point in time. This spatial signal gives information about the temperature differences between the various locations and may be useful to find the coldest or the hottest location. 
Another example of a multivariate signal is a digital movie, where each pixel represents the measured light intensity of a sensor in the video camera across time.
Multivariate signals are also common in various biomedical imaging applications, such as electroencephalography (EEG), magnetoencephalography (MEG), or magnetic resonance imaging (MRI). This book chapter will focus primarily on multivariate EEG signals.

Unlike time or spatial signals, multivariate signals can capture both temporal and spatial characteristics of the underlying system. The temporal characteristics can be analysed with conventional signal processing. For example, each signal can be analysed independently and the results can be collated across the signals. On the other hand, the spatial characteristics can be analysed using \textit{spectral graph theory}. 
This rests on the assumption that the spatially separated sensors are not fully independent of each other, but that the sensors are pairwise related to each other, forming a \textit{connectivity structure}. The goal of spectral graph theory is to analyse this connectivity structure \cite{atasoy2016human}.
However, neither strategy analyses the temporal and spatial characteristics jointly, which possibly disregards valuable interactions between the two.

Recently, GSP has emerged as a new field that promises to analyse the signals in time in dependence on their graph structure \cite{ortega2018graph,dong2020graph}. The principal idea behind many GSP-tools is to transform the data across space based on the sensor connectivity structure, before processing the signals with conventional tools. GSP has been successfully employed for neurophysiological imaging, such as functional MRI (fMRI) \cite{menoret2017evaluating, huang2018graph_perspective, itani2021graph, behjat2020spectral} or EEG \cite{mortaheb2019graph, saboksayr2021eeg, glomb2020connectome}. 
In the following, we illustrate different ways how GSP can be used for a variety of tasks, with a focus on applications for neurphysiological imaging.

\subsubsection{Graph Fourier transform (GFT) and dimensionality reduction}
In spectral graph theory, the graph is decomposed into \textit{graph eigenmodes}, which form the basis for the GFT.
Specifically, the GFT projects a spatial or a multivariate signal onto these eigenmodes. The projections are called the \textit{graph frequency signals}. Unlike the time signals, the graph frequency signals can be ordered. This renders it possible to reduce the dimensionality of the signal by selecting specific graph frequencies, e.g. the $k$ lowest or highest graph frequencies, depending on the problem. This method was used by Ménoret \textit{et al.} on a fMRI classification task \cite{menoret2017evaluating}. 
It turns out that this method is not only comparable to, but even equivalent to \textit{principal component analysis} (PCA) for certain graphs. This link is the topic of subsection \ref{ssec:link_pca}.
The role that this ordering plays for the graph frequency signals is the topic of the experimental part of this book chapter.

\subsubsection{Analysis of total variation}
The \textit{total variation} (TV) is a statistic which measures how much a signal varies on the graph structure. Subsection \ref{ssec:total_variation} derives two commonly used definitions for the total variation. While the previously examined spatial low-pass filter aims to reduce the TV of a signal on a graph, simply analysing the TV can also give useful insight about a signal. For example, Mortaheb \textit{et al}. found a correlation between the level of consciousness and the TV \cite{mortaheb2019graph}. Specifically, the authors analysed alpha-band signals in EEGs on a graph derived from the location of the sensors, giving insight into how disorders of consciousness affect communication in the brain relative to their range.

\subsubsection{Graph denoising}
GSP can be used to denoise a multivariate signal by applying a \textit{graph filter}. The underlying assumption of this tool is that closely connected sensors measure similar values. This assumption makes sense in the temperature measurement example: Temperature varies smoothly across geographic areas, meaning that nearby sensors measure similar values. GSP then allows to design a graph filter based on the connectivity structure, which reduces the variation between closely connected sensors. How such a filter can be constructed will be explained in subsection \ref{ssec:graph_signal_filtering}. Huang \textit{et al}. showcase how the method can be used both as a graph low-pass and as a graph high-pass filter \cite{huang2018graph_perspective}. The graph-filtered signal can then be used for further signal processing tasks.
This tool is the graph-equivalent of a low-pass filter in the temporal domain, which lowers the variation between nearby points in time.

\subsubsection{Graph learning}
Similar to the graph denoising tool, the \textit{graph learning} tool exploits the link between the graph structure and the spatial variation of a signal, i.d. the notion that a signal does not vary much on the graph structure.
However, instead of changing the signal to reduce the TV, this tool inversely changes the graph structure to reduce the TV. In other words, the TV is used as an optimisable objective function with the graph structure as its argument. Other terms can be added to this objective function to add further constraints on the graph. For example, one can further require the graph to have positive weights and to be sparse, i.d. to have few non-zero weights. 
While this graph-learning method has connections to learning Markov random fields, GSP adds a new perspective in terms of signals by interpreting the graph learning as minimising the TV \cite{dong2019learning}.
Kalofolias has introduced an algorithm in 2016 to solve this graph learning problem \cite{kalofolias2016learn}, which has since been picked up to retrieve graphs from neurophysiological signals \cite{menoret2017evaluating,mathur2022graph}. Section \ref{ssec:link_TV_FC} explores the link between this TV-based graph retrieval and functional connectivity-based graph retrieval.

\indent

Further applications of GSP include \textit{graph sampling}, which is an alternative way to reduce the spatial dimensionality of the multivariate signal, \textit{graph convolution}, which allows to pool features from nearby nodes and can be used for graph convolutional neural networks, and \textit{graph-based image compression} \cite{fracastoro2019graph}.

The theoretical section of this book chapter is intended to give a comprehensive introduction to GSP, specifically aiming to base the GSP concepts on concepts in classical signal processing, relate the various concepts to each other and explore the possible links between some of the concepts. Section \ref{sec:graphs_biomedical} will give preliminaries about graphs in general and graphs in biomedical imaging. The following section \ref{sec:GSP} will introduce GSP and some of its fundamental concepts in detail. It will further explore the similarities between graph retrieval methods and between graph representations for GSP and show how GSP can be linked to the PCA.

A considerable challenge for GSP is the plethora of possible graphs and graph representations that can be used to compute the GFT \cite{thanou2017learning}.
To begin with, there are several ways to retrieve the graph structure for the data. These can be data-driven, be acquired with a separate measurement or use knowledge about the data acquisition system (see section
\ref{ssec:graph_retrieval}). 
For some of the retrieved raw graphs, further preprocessing steps may be possible or required depending on the application, such as setting non-negative weights to zero, taking the absolute value of the weights, or normalising the graph.
Lastly, several graph representations can be computed from the preprocessed graphs, such as the adjacency matrix, the Laplacian matrix or the normalised Laplacian matrix.
While the type of graph retrieval and its representation can be relevant for the interpretation of the results in some cases, overall the GFT remains highly ambiguous. To counteract this ambiguity, suitable validation procedures are needed. Specifically, those validation procedures have to be able to isolate the performance gain induced by the graph structure from the performance of the method itself.
However, validation procedures for some recent applications of GSP have fallen short of pinpointing the superior performance to the graph structure.
A common neglect is the use of non-GSP models as baseline to validate GSP methods \cite{menoret2017evaluating}, which cannot validate the significance of the graph in the graph-based approach. A second neglect is the use of only one baseline model \cite{huang2018graph_perspective}, which may contain an unknown flaw and thereby fails to validate the method.
To sufficiently validate the role of the graph, we advocate the use of a baseline framework with multiple, GSP-based baseline models, which is the topic of subsection \ref{ssec:baseline_models}.
Our argument to use several baseline models is substantiated by our experimental results (see section \ref{sec:results}), where two baseline models outperform a third baseline model.

One goal of the experimental section of this book chapter is to evaluate the applicability of GSP to neurophysiological signals. The general idea is to reduce the dimensionality of the signal analysis by finding a selection of graph frequency signals which can be used to analyse the signal instead of the full multivariate signal. Specifically, the quality of spectral features in individual graph frequency signals is assessed by evaluating their usefulness on a classification task. 
Here, the scarcity of the available real data is a limitation, as each independent signal corresponds to one measurement of a patient. To overcome this limitation, we developed a simulation algorithm to generate arbitrary amounts of data, which is described in subsection \ref{ssec:simulation}. 
The simulated signals are equipped with a temporal and a spatial structure. While developed for the problem at hand, the simulated data may be useful for other problems in biomedical imaging as well.
Due to its unlimited size, the artificial data set allows to analyse the graph frequency signals individually, instead of having to rely on group averages. This enables us to explore the individual role that each of the graph frequencies plays. 
The methodology is explained in section \ref{sec:methodology}. The results, shown in section \ref{sec:results}, suggest that the higher graph frequency signals may be more useful for data classification than the raw time signals, while falling short of attributing this performance boost to the use of the connectivity structure.

\section{Graphs in biomedical imaging}
\label{sec:graphs_biomedical}

\subsection{Graphs}
\label{ssec:brainnetworks}

Formally, networks can be represented by a \textit{weighted graph} $\mathcal{G}\coloneqq (\mathcal{V}, \mathbf{A})$, which consists of $N$ vertices $\mathcal{V}=\{1, 2, 3, ..., N\}$ and the \textit{weighted adjacency matrix} $\mathbf{A}\in \mathbb{R}^{N\times N}$. The entries $a_{ij}$ of $\mathbf{A}$ represent the strength of the connectivity between node $i$ and node $j$. 
If the connectivities between each pair of nodes are symmetric, or $a_{ij}=a_{ji}$ for any $i$ and $j$, then $\mathbf{A}$ is symmetric and the graph is said to be undirected; conversely, the graph is said to be directed. Furthermore, if the diagonal elements $a_{ii}$ are all zero, meaning that the nodes are not connected to themselves with loops, the graph is called a \textit{simple graph}. Figure \ref{fig:graphs}A depicts a weighted directed simple graph with $N=6$ vertices.

A second, useful representation of a graph $\mathcal{G}$ is its \textit{Laplacian matrix}. The Laplacian can be directly computed from the adjacency matrix $\mathbf{A}$ as $\mathbf{L} = \mathbf{D} - \mathbf{A}$. Here, $\mathbf{D}=\mathrm{diag}(\mathbf{A}\cdot \mathbf{1})$ denotes the \textit{degree matrix}.
Alternatively, the symmetric normalised Laplacian can be used instead of the Laplacian, which is computed as $\mathbf{L}_\mathrm{norm}=\mathbf{D}^{-1/2}\mathbf{L}\mathbf{D}^{-1/2}$.
While the weighted adjacency matrix can be viewed as a graph shift operator (GSO, section \ref{ssec:GSO}), the Laplacian can be associated with the negative difference operator (section \ref{ssec:Laplacian_as_diff}).

\subsection{Graph shift operator (GSO)}
\label{ssec:GSO}
The weighted adjacency matrix is the canonical algebraic representation of a graph. This subsection further explores the role of the adjacency matrix as a GSO and its implications for graph dynamics. In particular, these graph dynamics are capitalised for generating the artificial multivariate signals in subsection \ref{ssec:simulation}.

\begin{figure}[tbh]
    \centering
    \resizebox{\columnwidth}{!}{%


\renewcommand*\familydefault{\sfdefault} 
\fontfamily{\sfdefault}\selectfont

\noindent
\begin{tikzpicture}[bend angle=10]
  \draw (-0.5, 1.15) node {A};
  \node[draw, circle, fill=dodgerblue!20, thick, minimum size=0.5cm] (t0) at (1.5,0){$1$};
  \node[draw, circle, fill=dodgerblue!20, thick, minimum size=0.5cm,shift={(1.6, 1.1)}] (t1) at (t0){$2$};
  \node[draw, circle, fill=dodgerblue!20, thick, minimum size=0.5cm,shift={(2.1, -0.8)}] (t2) at (t0){$3$};
  \node[draw, circle, fill=dodgerblue!20, thick, minimum size=0.5cm,shift={(3.2, 0)}] (t3) at (t0){$4$};
  \node[draw, circle, fill=dodgerblue!20, thick, minimum size=0.5cm,shift={(4.2, 0.9)}] (t4) at (t0){$5$};
  \node[draw, circle, fill=dodgerblue!20, thick, minimum size=0.5cm,shift={(4.6,-0.6)}] (t5) at (t0){$6$};

  \draw[-latex,thick,bend right,line width=0.1mm] (t2) edge (t1) node[text width=0.8cm,align=center, font={\tiny}, midway,below] {};
  \draw[-latex,thick,bend right,line width=0.3mm] (t1) edge (t2) node[text width=0.8cm,align=center, font={\tiny}, midway,below] {};
  \draw[-latex,thick,bend right,line width=0.35mm] (t0) edge (t1) node[text width=0.8cm,align=center, font={\tiny}, midway,below] {};
  \draw[-latex,thick,bend right,line width=0.5mm] (t1) edge (t0) node[text width=0.8cm,align=center, font={\tiny}, midway,below] {};
  \draw[-latex,thick,line width=0.15mm] (t0) -- (t2) node[text width=0.8cm,align=center, font={\tiny}, midway,below] {};
  \draw[-latex,thick,line width=0.4mm] (t3) -- (t0) node[text width=0.8cm,align=center, font={\tiny}, midway,below] {};
  \draw[-latex,thick,line width=0.2mm] (t3) -- (t4) node[text width=0.8cm,align=center, font={\tiny}, midway,below] {};
  \draw[-latex,thick,bend right,line width=0.2mm] (t1) edge (t3) node[text width=0.8cm,align=center, font={\tiny}, midway,below] {};
  \draw[-latex,thick,bend right,line width=0.15mm] (t3) edge (t1) node[text width=0.8cm,align=center, font={\tiny}, midway,below] {};
  \draw[-latex,thick,bend right,line width=0.35mm] (t3) edge (t5) node[text width=0.8cm,align=center, font={\tiny}, midway,below] {};
  \draw[-latex,thick,bend right,line width=0.45mm] (t5) edge (t3) node[text width=0.8cm,align=center, font={\tiny}, midway,below] {};

  \draw (-0.5, -1.5) node {B};
  \node[draw, circle, fill=dodgerblue!20, thick, minimum size=0.5cm] (t0) at (0,-2.8){$t_0$};
  \node[draw, circle, fill=dodgerblue!20, thick, minimum size=0.5cm, shift={(0.8,0,0)}] (t1) at (t0.east){$t_1$};
  \node[draw, circle, fill=dodgerblue!20, thick, minimum size=0.5cm, shift={(0.8,0,0)}] (t2) at (t1.east){$t_2$};
  \node[draw, circle, fill=dodgerblue!20, thick, minimum size=0.5cm, shift={(0.8,0,0)}] (t3) at (t2.east){$t_3$};
  \node[draw, circle, fill=dodgerblue!20, thick, minimum size=0.5cm, shift={(0.8,0,0)}] (t4) at (t3.east){$t_4$};
  \node[draw, circle, color=white!0, fill=white!0, thick, minimum size=0.5cm, shift={(0.8,0,0)}] (t_inv1) at (t4.east){};
  \node[draw, circle, color=white!0, fill=white!0, thick, minimum size=0.5cm, shift={(0.8,0,0)}] (t_inv2) at (t_inv1.east){};
  
  \node[circle,fill,inner sep=1.2pt, shift={(0.98,0)}] at (t4.east){};
  \node[circle,fill,inner sep=1.2pt, shift={(1.33,0)}] at (t4.east){};
  \node[circle,fill,inner sep=1.2pt, shift={(1.68,0)}] at (t4.east){};
  
  \node[draw, circle, fill=dodgerblue!20, thick, minimum size=0.5cm, shift={(0.8,0,0)}] (t99) at (t_inv2.east){\phantom{$t_0$}};
  \node[shift={(0.8,0,0)}] (t99_text) at (t_inv2.east){$\,t$\tiny $_{N-1}$};

  \draw[-latex,thick] (t0.east) -- (t1.west) node[text width=0.8cm,align=center, font={\tiny}, midway,below] {};
  \draw[-latex,thick] (t1.east) -- (t2.west) node[text width=0.8cm,align=center, font={\tiny}, midway,below] {};
  \draw[-latex,thick] (t2.east) -- (t3.west) node[text width=0.8cm,align=center, font={\tiny}, midway,below] {};
  \draw[-latex,thick] (t3.east) -- (t4.west) node[text width=0.8cm,align=center, font={\tiny}, midway,below] {};
  \draw[-latex,thick] (t4.east) -- (t_inv1.west) node[text width=0.8cm,align=center, font={\tiny}, midway,below] {};

  \draw[-latex,thick] (t_inv2.east) -- (t99.west) node[text width=0.8cm,align=center, font={\tiny}, midway,below] {};
  
  \draw[-latex,thick] (t99.north west) to [out=158,in=22] (t0.north east);

\end{tikzpicture}%
    }
    \vspace{1em}
    \caption{(A) Graph with six nodes, which can be represented by an adjacency matrix $\mathbf{A}$. The displayed connections are directed, meaning that they can go in both directions. They are also weighted, which is indicated by the line width of the arrows. The graph can represent an arbitrary spatial graph topology of a multivariate signal acquired from a measurement system with six sensors. (B) Time graph with linear topology. The graph connects each time step $t_i$ to the subsequent time step $t_{i+1}$, thereby shifting a signal in time. For periodic signals, the last time step is connected to the first time step. The graph can be represented by the cyclic shift matrix $\mathbf{A}_c$}
    \label{fig:graphs}
\end{figure}
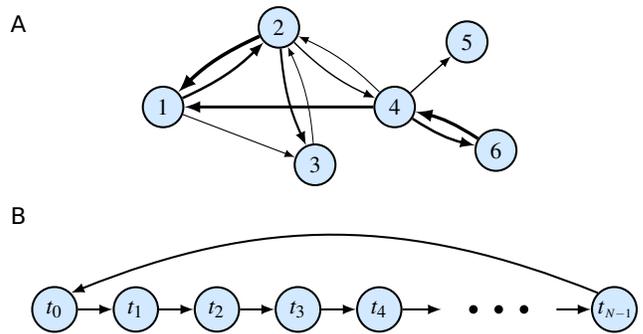

In discrete signal processing (DSP), a cyclic time signal $s$ with $N$ samples can be represented algebraically by a time-ordered vector $\mathbf{s}=(s_0,...,s_{N-1})^\top$, where $s_0$ and $s_{N-1}$ are the time samples at the first and the last time step, respectively. The shift operator $T_1$ shifts each time sample $s_n$ to the subsequent time step,
\begin{align}
    T_1: (s_0,...,s_{N-1})^\top \mapsto (s_{N-1},s_0,...,s_{N-2})^\top,
\end{align}
and can be algebraically represented by the cyclic shift matrix
\begin{align}
\label{eq:cyclic_shift}
    \mathbf{A}_c = \begin{bmatrix} 
    0 & 0 & 0 & \dots  & 1\\
    1 & 0 & 0 & \dots  & 0\\
    0 & 1 & 0 & \dots  & 0\\
    \vdots && \ddots & \ddots & \vdots\\
    0 & & \dots & 1  & 0 
    \end{bmatrix}.
\end{align}
Note that the last time sample $s_{N-1}$ is shifted to the first time step, as we assume the time signal to be cyclic. 

Importantly, $\mathbf{A}_c$ can be interpreted as the adjacency matrix of a graph with a linear topology with $N$ vertices, as shown in Figure \ref{fig:graphs}B.
The interpretation of the discretised time dimension as a linear time graph allows to generalise the shift operator to more arbitrary graph topologies, such as the one shown in Figure \ref{fig:graphs}A. Consequently, the GSO of a graph is simply given by its weighted adjacency matrix $\mathbf{A}$. It spatially shifts a signal at node $i$ to its neighbouring nodes, weighted by the strength of the connection. 
From a graph dynamics perspective, the GSO can be approximately thought of as shifting a signal at each node to its neighbouring nodes in each time step, due to the connections between the nodes. Hence, the GSO can be thought of as a ``time evolution operator'', as it is evolving the graph signal in time. This approximation is very crude and only describes the dynamics of the signals which are due to the connectivity structure. Nevertheless, in subsection \ref{ssec:simulation} this property of the GSO proves to be useful to simulate a mutivariate signal.
A more sophisticated way to ascribe the graph dynamics to the connectivity structure is to model the dynamics as a heat diffusion process \cite{thanou2017learning}, which involves the Laplacian operator. The Laplacian operator itself is based on the GSO, as shown in the following section \ref{ssec:Laplacian_as_diff}.

\subsection{Laplacian as negative difference operator}
\label{ssec:Laplacian_as_diff}
In discrete calculus, the operator $\Delta$ defines a forward difference in terms of the shift operator $T_1$ and the identity operator $I$:
\begin{align}
    \Delta = T_1 - I.
\end{align}
In the case of a finite, cyclic signal, the shift operator can be represented algebraically by the cyclic shift matrix $\mathbf{A}_c$ as defined in (\ref{eq:cyclic_shift}), while the identity operator can be represented by the degree matrix $\mathbf{D}_c=\mathbbm{1}$. The negative forward difference operator $-\Delta$ can then be represented by the Laplacian matrix $\mathbf{L}_c$:
\begin{align}
    \mathbf{L}_c = \mathbf{D}_c - \mathbf{A}_c.
\end{align}

Likewise, the second-order central difference operator $\Delta^2$ is given by:
\begin{align}
    \Delta^2 = T_1 - 2I + T_{-1},
\end{align}
where $T_{-1}$ is the right shift operator:
\begin{align}
    T_{-1}: (s_0,...,s_{N-1})^\top \mapsto (s_1,...,s_{N-1},s_0)^\top.
\end{align}
As $T_{-1}$ can be represented algebraically by the transposed adjacency matrix $\mathbf{A}_c^\top$, the symmetrised Laplacian matrix,
\begin{align}
    \mathbf{L}_{c,\mathrm{sym}} = \mathbf{D}_c - \frac{\left(\mathbf{A}_c + \mathbf{A}_c^\top\right)}{2},
\end{align}
 is the algebraic representation of the operator $-2 \Delta^2$.

\subsection{Multivariate signals in biomedical imaging}
Biomedical imaging is one of many applications where multivariate signals are acquired.
Brain imaging techniques, such as fMRI, EEG, or MEG, record multiple signals simultaneously in time at different spatial locations. In fMRI, the signal locations are called \textit{voxels}, whereas in MEG and EEG they are called \textit{channels}. An EEG measurement setup with $N_c$ channels which record the brain over a time period $T$ at a sampling rate $f$ yields a data matrix $\mathbf{X}\in \mathbb{R}^{N_c\times N_t}$, where $N_t=\lfloor f\cdot T\rfloor$ is the number of time samples. The $i$-th row of $\mathbf{X}$ correspond to the \textit{temporal signal}, or simply signal, at location $i$ and can be denoted as $\mathbf{x}_{i*}$. On the other hand, the $j$-th column corresponds to the spatial, or \textit{graph signal} at time sample $j$, which we denote as $\mathbf{x}_{*j}$.

\subsection{Graph retrieval}
\label{ssec:graph_retrieval}
The GFT of the multivariate signals relies on the graph, or spatial structure, of the data. 
The retrieval of the graph from neuroimaging data is not unique, but can be based on the structural, or anatomical, connectivity, the functional connectivity \cite{friston1993functional}, or the geometric location of the sensors \cite{menoret2017evaluating}.

Firstly, the graph can be based on the \textit{functional connectivity} between the signals.
This method is purely data-driven and can therefore be used on all data sets.
Common choices to build the graph include computing pairwise Pearson correlations or covariances, but nonlinear measures such as mutual information, the phase lage index, or the phase locking value can be used as well \cite{he2021nonlinear}.
As shown in subsection \ref{ssec:link_pca}, there is a link between GFT using functionally retrieved graphs and PCA.
The experimental section of this chapter uses the Pearson correlation to construct the graph (see subsection \ref{ssec:analysis}).
Secondly, a graph can be constructed using the \textit{structural connectivity} between nodes, which can be determined with secondary measurements. For neurophysiological signals, for example, diffusion tensor imaging can be employed \cite{damoiseaux2009greater}.
Lastly, the connectivity between the nodes can be determined by their \textit{geometric} properties, such as the pairwise distances between them. The distances can then be mapped to the connectivity strengths. The method only requires knowledge about the data acquisition system.

In some instances, the graph retrieval methods can yield similar results. This is not surprising, because the methods are principally aiming to estimate the same connectivity structure underlying the neural substrate. For example, the structural and the functional connectivity are generally related to each other \cite{damoiseaux2009greater}. However, note that Wang \textit{et al.} give examples why the structural connectivity is not sufficient to explain the dynamics of neural activity \cite{wang2016brain}, and thus the functional connectivity.

Despite some similarities, Horwitz argued that there is no single underlying connectivity structure, but that connectivity should be thought of as ``forming a class of concepts with multiple members'' \cite{horwitz2003elusive}. Generally, the choice of the graph retrieval method can alter the interpretation of the graph \cite{fraschini2019robustness}. 
As an example of this, Mortaheb \textit{et al.} \cite{mortaheb2019graph} used a geometric graph to calculate the edge-based total variation, which allowed them to interpret the results as effects of local communication.
One important difference between the graph retrieval methods lies in whether they allow negative connections between nodes or not. For example, some functional connectivity measures, such as the pairwise Pearson correlation, can retrieve negative connections. On the other hand, structural connectivity measures looking at physical structures in the brain, such as white matter projections, typically cannot discern whether a connection is positive or negative \cite{sporns2022structure}. The implications of using negative weights in the adjacency matrix are further explored in subsection \ref{ssec:negative}.

\section{Graph signal processing (GSP)}
\label{sec:GSP}
\begin{figure*}[tbh]
    \centering
    \resizebox{0.8\textwidth}{!}{%
    \renewcommand*\familydefault{\sfdefault}
\fontfamily{\sfdefault}\selectfont

\noindent
\begin{tikzpicture}
  \node[draw, fill=lightwheat, align=center, font={\smaller}] (SO) at (0, 0){shift \\operator};
  \node[draw, fill=lightwheat, align=center, font={\smaller}, shift={(0.75, -1.434)}] (TV) at (SO){TV};
  \node[draw, fill=lightwheat, align=center, font={\smaller}, shift={(-3.0, -3.0)}] (deriv_DFT) at (SO){derivative-based\\interpretation\\DFT};
  \node[draw, fill=lightwheat, align=center, font={\smaller}, shift={(3.0, -3.0)}] (shift_DFT) at (SO){\phantom{derivative-based}\\\phantom{interpretation}\\\phantom{DFT}};
  \node[align=center, font={\smaller}, shift={(3.0, -3.0)}] (shift_DFT_text) at (SO){filter-based\\interpretation\\DFT};
  \node[draw, fill=lightwheat, align=center, font={\smaller}, shift={(0, -4.25)}] (conv) at (SO){convo-\\lution};
  \node[draw, fill=lightwheat, align=center, font={\smaller}, shift={(0, -6.0)}] (filter) at (SO){digital\\filter};
  \node[draw, fill=lightwheat, align=center, font={\smaller}, shift={(0, -7.5)}] (spectral_wavelets) at (SO){spectral\\wavelets};

  \draw[-latex,thick] ($(SO.south west)!0.75!(SO.south east)$) -- (TV.north);
  \draw[-latex,thick] (SO.south west) -- (deriv_DFT.north) node[minimum size=0.0cm,align=center, midway](shift2derivDFT) {};
  \draw[-latex,thick] (SO.south east) -- (shift_DFT.north) node[minimum size=0.0cm,align=center,midway](shift2shiftDFT) {};
  \draw[-latex,thick] (TV.west) -- (shift2derivDFT);
  \draw[-latex,thick] (TV.east) -- (shift2shiftDFT);
  \draw[-latex,thick] (SO.south) -- (conv.north);
  \draw[-latex,thick] (conv.south) -- (filter.north) node[text width=0.0cm,align=center,midway,below](conv2filt) {};
  \draw[latex-latex,thick] (deriv_DFT.east) -- (shift_DFT.west);
  \draw[-,thick,bend left=0] (shift_DFT.south) edge[-latex] (conv2filt);
  \draw[-,thick,bend left,dashed] ($(shift_DFT.south west)!0.75!(shift_DFT.south east)$) edge[-latex,out=-8,in=195] (filter.north east);
  \draw[-latex,thick] (filter.south) -- (spectral_wavelets.north);
  
  \node[draw, fill=lightwheat, align=center, font={\smaller}, shift={(9.5, 0)}] (GSO) at (SO){graph\\shift operator};
  \node[draw, fill=lightwheat, text width=0.8cm, align=center, font={\smaller}, shift={(-0.75, -1.425)}] (deriv_TV) at (GSO){edge\\TV};
  \node[draw, fill=lightwheat, text width=0.8cm, align=center, font={\smaller}, shift={(0.75, -1.425)}] (shift_TV) at (GSO){node\\TV};
  \node[draw, fill=lightwheat, align=center, font={\smaller}, shift={(-3.0, -3.0)}] (deriv_GFT) at (GSO){derivative-based\\GFT\\(Laplacian)};
  \node[draw, fill=lightwheat, align=center, font={\smaller}, shift={(3.0, -3.0)}] (shift_GFT) at (GSO){\phantom{derivative-based}\\\phantom{GFT}\\\phantom{(Laplacian)}};
  \node[align=center, font={\smaller}, shift={(3.0, -3.0)}] (shift_GFT_text) at (GSO){filter-based\\GFT\\(adjacency)};
  \node[draw, fill=lightwheat, align=center, font={\smaller}, shift={(0, -4.5)}] (graph_conv) at (GSO){graph\\convolution};
  \node[draw, fill=lightwheat, align=center, font={\smaller}, shift={(3.0, -6.0)}] (graph_filter) at (GSO){\phantom{graph filter}\\\phantom{(adjacency)}};
  \node[align=center, font={\smaller}, shift={(3.0, -6.0)}] (graph_filter_text) at (GSO){graph filter\\(adjacency)};
  \node[draw, fill=lightwheat, align=center, font={\smaller}, shift={(-3.0, -6.0)}] (graph_filter_alt) at (GSO){graph filter\\(Laplacian)};
  
  \node[draw, fill=lightwheat, align=center, font={\smaller}, shift={(3.0, -7.5)}] (graph_spectral_wavelets) at (GSO){\phantom{graph wavelets}\\\phantom{(adjacency)}};
  \node[align=center, font={\smaller}, shift={(3.0, -7.5)}] (graph_spectral_wavelets_text) at (GSO){graph wavelets\\(adjacency)};
  \node[draw, fill=lightwheat, align=center, font={\smaller}, shift={(-3.0, -7.5)}] (graph_spectral_wavelets_alt) at (GSO){graph wavelets\\(Laplacian)};

  \draw[-latex,thick] ($(GSO.south west)!0.25!(GSO.south east)$) -- (deriv_TV.north);
  \draw[-latex,thick] ($(GSO.south west)!0.75!(GSO.south east)$) -- (shift_TV.north);
  \draw[-latex,thick] (GSO.south west) -- (deriv_GFT.north) node[minimum size=0.0cm,align=center, font={\tiny}, midway](shift2derivGFT) {};
  \draw[-latex,thick] (GSO.south east) -- (shift_GFT.north) node[minimum size=0.0cm,align=center, font={\tiny}, midway](shift2shiftGFT) {};
  \draw[-latex,thick] (deriv_TV.west) -- (shift2derivGFT);
  \draw[-latex,thick] (shift_TV.east) -- (shift2shiftGFT);
  \draw[-latex,thick] (GSO.south) -- (graph_conv.north);

  \node[text width=0.0cm, align=center, shift={(0, -0.6)}] (TVs_midway) at ($(deriv_TV.south)!0.5!(shift_TV.south)$,0){};
  \draw[-,thick,bend right=50] (deriv_TV.south) edge[latex-latex] (shift_TV.south) {};
  \draw[red,thick,shift={($(deriv_TV.south)!0.5!(TVs_midway)$,0)}](-0.1,-0.1)--(0.1,+0.1);
  \draw[red,thick,shift={($(deriv_TV.south)!0.5!(TVs_midway)$,0)}](-0.1,+0.1)--(0.1,-0.1);
  \draw[red,thick,shift={($(TVs_midway)!0.5!(shift_TV.south)$,0)}](-0.1,-0.1)--(0.1,+0.1);
  \draw[red,thick,shift={($(TVs_midway)!0.5!(shift_TV.south)$,0)}](-0.1,+0.1)--(0.1,-0.1);
  
  \draw[latex-latex,thick] (deriv_GFT.east) -- (shift_GFT.west) node[midway](GFTs_midway) {};
  
  \draw[red,thick,shift={($(deriv_GFT.east)!0.4!(GFTs_midway)$,0)}](-0.1,-0.1)--(0.1,+0.1);
  \draw[red,thick,shift={($(deriv_GFT.east)!0.4!(GFTs_midway)$,0)}](-0.1,+0.1)--(0.1,-0.1);
  \draw[red,thick,shift={($(deriv_GFT.east)!0.8!(GFTs_midway)$,0)}](-0.1,+0.1)--(0.1,-0.1);
  \draw[red,thick,shift={($(deriv_GFT.east)!0.8!(GFTs_midway)$,0)}](-0.1,-0.1)--(0.1,+0.1);
  \draw[red,thick,shift={($(GFTs_midway)!0.2!(shift_GFT.west)$,0)}](-0.1,+0.1)--(0.1,-0.1);
  \draw[red,thick,shift={($(GFTs_midway)!0.2!(shift_GFT.west)$,0)}](-0.1,-0.1)--(0.1,+0.1);
  \draw[red,thick,shift={($(GFTs_midway)!0.6!(shift_GFT.west)$,0)}](-0.1,+0.1)--(0.1,-0.1);
  \draw[red,thick,shift={($(GFTs_midway)!0.6!(shift_GFT.west)$,0)}](-0.1,-0.1)--(0.1,+0.1);

  \draw[-latex,thick] (graph_conv.south east) -- (graph_filter.north west) node[midway](conv2filt) {};
  \draw[-latex,thick] ($(shift_GFT.south west)!0.25!(shift_GFT.south east)$) -- (conv2filt);
  \draw[-,thick,bend left=-25,dashed] ($(shift_GFT.south west)!0.375!(shift_GFT.south east)$) edge[-latex] ($(graph_filter.north west)!0.25!(graph_filter.north east)$);
  \draw[-latex,thick,dashed] (deriv_GFT.south) -- (graph_filter_alt.north) node[midway](deriv_GFT2graph_filter_alt) {};
  \draw[-latex,thick] (graph_conv.south west) -- (graph_filter_alt.north east) node[midway](conv2filt_alt) {};
  
  \draw[-latex,thick] (graph_filter.south) -- (graph_spectral_wavelets.north);
  \draw[-latex,thick] (graph_filter_alt.south) -- (graph_spectral_wavelets_alt.north);
  
  \draw[red,thick,shift={(conv2filt_alt)}](-0.1,-0.1)--(0.1,+0.1);
  \draw[red,thick,shift={(conv2filt_alt)}](-0.1,+0.1)--(0.1,-0.1);


  \draw[navyblue_mod,-latex,thick] (SO.east) -- (GSO.west) node[navyblue_mod,minimum size=0.5cm,align=center, font={\small}, midway, below] {graph extension};

  \draw[navyblue_mod,-latex,thick] (1.8, -4.58) -- (deriv_GFT2graph_filter_alt) node[navyblue_mod,minimum size=0.5cm,align=center, font={\small}, midway, below] {by analogy};
\end{tikzpicture}
    }
    \caption{Interconnections between discrete signal processing (DSP) concepts and its graph extensions. In DSP, the DFT can be traced back to the shift operator. Crucially, a derivative-based interpretation and a filter-based interpretation of the DFT are equivalent. The TV increases with increasing Fourier frequency, thereby linking the two concepts. Convolution, digital filters and finally spectral wavelets are built on the concept of the shift operator, whereby the digital filter can be expressed using the DFT. Extending the shift operator to graphs allows to build analogous concepts for graphs. Importantly, the derivative- and the filter-based GFT are not equivalent. The graph Fourier modes in both GFTs can be ordered by their frequency using either the edge-based or the node-based TV. The graph convolution can be used to define an adjacency matrix-based graph filter. The similar, Laplacian matrix-based graph filter can only be constructed by analogy and is not directly linked to the graph convolution. Graph spectral wavelets are derived from the graph convolution}
    \label{fig:high_level_connections}
\end{figure*}
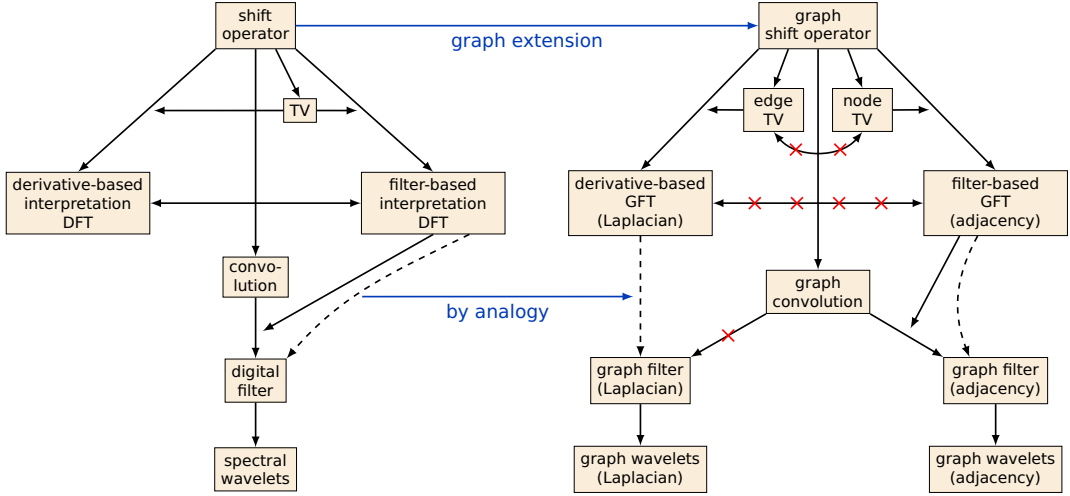

In this chapter, GSP concepts are developed in analogy to classical signal processing. Figure \ref{fig:high_level_connections} gives a comprehensive overview of the relations between the relevant concepts in classical signal processing and how they are extended to GSP.
Importantly, this extension is ambiguous and leads to two separate definitions of the GFT, from which several other concepts are derived. 

\subsection{Graph Fourier Transform (GFT)}
\label{ssec:GFT}
The GFT is an extension of the Fourier transform to graphs and is at the heart of GSP. Here, we show that it can be derived either via graph filtering (\textit{filter-based GFT}) \cite{sandryhaila2014discrete,ortega2018graph} or via discrete derivatives on graphs (\textit{derivative-based GFT}) \cite{hammond2011wavelets,sandryhaila2014discrete}. Both these approaches are built on the extension of the shift operator to graphs, i.e., the GSO given by the adjacency matrix $\mathbf{A}$. While both approaches yield the same result in DSP, their extensions to GSP turn out to be slightly different: The derivative-based GFT is composed of the eigenvectors of the adjacency matrix, whereas the filter-based GFT is composed of the eigenvectors of the Laplacian matrix.

The two approaches mirror the two definitions for the total variation in section \ref{ssec:total_variation}.
Section \ref{ssec:rel_adj_lap} shows theoretically why the two approaches are nevertheless similar to each other in some practical applications.

\subsubsection{Filter-based Graph Fourier Transform}
\label{sssec:filter_GFT}
The derivation in this subsection follows that of Ortega \textit{et al.} \cite{ortega2018graph}. Let s be a time signal represented by a time-ordered vector $\mathbf{s}=(s_0,...,s_{N-1})^\top$ and $T_1$ a shift operator $\mathbf{A}_c$ represented algebraically by $\mathbf{A}_c$, as defined in equation (\ref{eq:cyclic_shift}).
A finite impulse response filter $h$ of order $N$ can be characterised as a sum of the $N$ samples $s_{in}[n-i \mod N]$ preceding the time step $n$, weighted by $p_i$:
\begin{align}
    s_{out}[n] = (h \cdot s_{in})[n] = \sum_{i=0}^{N-1}p_i s_{in}[n-i \mod N].
\end{align}

Importantly, this means that the filter $h$ can be represented as a $N$-th order polynomial in the time shift operator $T_1$, as defined in subsection \ref{ssec:GSO}:
\begin{align}
    h = \sum_{i=0}^{N-1}p_i T_1^i.
\end{align}
Algebraically, the filter is given by the matrix $\mathbf{H}$ as a polynomial in the cyclic shift matrix $\mathbf{A}_c$ as defined in equation (\ref{eq:cyclic_shift}):
\begin{align}
    \mathbf{H} = \sum_{i=0}^{N-1}p_i \mathbf{A}_c^i.
\end{align}

The eigenvectors of $\mathbf{A}_c$ are the complex exponentials $\mathbf{v}_k = (\omega^{0}, \omega^{k}, ..., \omega^{(N-1)k})^\top/\sqrt{N}$, with $\omega=e^{2\pi j / N}$ and the imaginary unit $j=\sqrt{-1}$. The eigenvalue for each complex exponential $\mathbf{v}_k$ is $\omega^k$. Consequently, the eigendecomposition of $\mathbf{A}_c$ is given by:
\begin{align}
\label{eq:Ac_eigendecomposition}
    \mathbf{A}_c &= \mathbf{Q}_c\Lambda_c\,\mathbf{Q}_c^{-1}\\&= (\mathbf{v}_1, \mathbf{v}_2, ..., \mathbf{v}_N)\begin{bmatrix}
    \omega^0 & & & \\
    & \omega^1 & & \\
    & & \ddots & \\
    & & & \omega^{N-1}
    \end{bmatrix}(\mathbf{v}_1^*, \mathbf{v}_2^*, ..., \mathbf{v}_N^*)^\top.
\end{align}
Crucially, the matrix $\mathbf{Q}_c^{-1}$, comprising the complex conjugated exponentials $\mathbf{v}_k$, can be identified as the discrete Fourier transform matrix $\mathbf{DFT}_N$, yielding:
\begin{align}
    \label{eq:Ac_DFT}
    \mathbf{A}_c = \mathbf{DFT}_N^{-1}\Lambda_c\,\mathbf{DFT}_N.
\end{align}

The eigenvectors $\mathbf{v}_k$ of $\mathbf{A}_c$ are also eigenvectors of the filter matrix $\mathbf{H}$,
\begin{align}
\label{eq:H_through_time}
    \mathbf{H}\mathbf{v}_k = \sum_{i=0}^{N-1}p_i \mathbf{A}_c^i \mathbf{v}_k = \left(\sum_{i=0}^{N-1}p_i \left(\omega^{k}\right)^i\right)\mathbf{v}_k,
\end{align}
which of course is the well-known fact in DSP that filters can change the magnitude or the phase of a sinusoidal signal, but not its frequency.

The central role that the shift operator plays in constructing the filter and its relation to the DFT justifies to define the GFT in terms of the GSO, which can be represented algebraically by the adjacency matrix $\mathbf{A}$ (see subsection \ref{ssec:GSO}). In analogy to DSP, the filter-based GFT for a spatial signal $\mathbf{x}\in \mathbb{R}^{N_c}$ on a graph $\mathbf{A}$ is defined as follows:
\begin{align}
    \label{eq:A_GFT}
    \mathbf{A} =&\; \mathbf{Q}\Lambda\mathbf{Q}^{-1} \\
    \mathbf{GFT}_\mathbf{A} \coloneqq&\; \mathbf{Q}^{-1}\\
    \label{eq:GFT_A}
    \mathbf{\tilde{x}} \coloneqq&\; \mathbf{GFT}_\mathbf{A} \mathbf{x},
\end{align}
where $\mathbf{\tilde{x}}$ is the transformed spatial signal. The eigenvalues $\lambda_k$ of $\mathbf{A}$ are the diagonal elements of $\Lambda$. If $\mathbf{A}$ is symmetric, they are real-valued and can be sorted in ascending order, and
the eigenvectors $\mathbf{v}_k$ become orthogonal. 

However, other matrices, such as the Laplacian matrix $\mathbf{L}_c = \mathbf{D}_c - \mathbf{A}_c$, have the same eigenvectors as $\mathbf{A}_c$ and can be similarly decomposed using the DFT-matrix. Importantly, this degeneracy in the DSP case is lifted when the cyclic graph is extended to more complex graph topologies, as also illustrated in the overview graphic \ref{fig:high_level_connections}. This yields the alternative, derivative-based definition of the GFT, which is derived in the following subsection.

\subsubsection{Derivative-based Graph Fourier Transform}
In the classical Fourier transform, Fourier modes are complex exponentials $e^{-j\omega t}$. These exponentials are trivially eigenfunctions of the partial time-derivative $\partial / \partial t$ \cite{hammond2011wavelets,sandryhaila2014discrete}, as well as the second partial time-derivative $\partial^2 / \partial t^2$,
\begin{align}
    \frac{\partial}{\partial t} e^{-j\omega t} &= -j\omega e^{-j\omega t}\\
    \frac{\partial^2}{\partial t^2} e^{-j\omega t} &= -\omega^2 e^{-j\omega t},
\end{align}
with eigenvalues of $-j\omega$ and $-\omega^2$, respectively.

As shown in subsection \ref{ssec:Laplacian_as_diff}, in DSP the directed Laplacian matrix $\mathbf{L}_c$ of a cyclic shift graph is the algebraic representation of the negated forward difference operator $-\Delta$, whereas the symmetrised Laplacian matrix $\mathbf{L}_{c,\mathrm{sym}}$ is the algebraic representation of the operator $-2\Delta^2$.
Crucially, the eigenvectors $\mathbf{v}_k$ of $\mathbf{A}_c$, which constitute the discrete Fourier modes, are also eigenvectors of $\mathbf{L}_c$ with eigenvalues $\lambda_k' = (1 - \lambda_k)$:
\begin{align}
    \mathbf{L}_c \mathbf{v}_k &= \left(\mathbbm{1} - \mathbf{A}_c\right)\mathbf{v}_k = 1 \cdot \mathbf{v}_k - \lambda_k \mathbf{v}_k = \lambda_k' \mathbf{v}_k\\
    \lambda_k' &= (1 - \lambda_k).
\end{align}
In other words, the discrete Fourier modes are eigenvectors of the negated forward difference operator, in the same way that continuous Fourier modes are eigenfunctions of the time-derivative.
Finally, the eigendecomposition of $\mathbf{L}_c$ is given by:
\begin{align}
    \mathbf{L}_c = \mathbf{DFT}_N^{-1}(\mathbbm{1} - \Lambda_c)\,\mathbf{DFT}_N \eqqcolon \mathbf{DFT}_N^{-1}\Lambda'_c\,\mathbf{DFT}_N,
\end{align}
where $\Lambda_c$ is the eigenvalue matrix of $\mathbf{A}_c$ as given in (\ref{eq:Ac_DFT}).

Interpreting Fourier modes as eigenmodes of the derivative and discrete Fourier modes as eigenmodes of the forward difference leads to the second extension of DFT to graphs. Accordingly, given the adjacency matrix $\mathbf{A}$ of a graph, the derivative-based GFT is defined as follows:
\begin{align}
    \mathbf{L} =&\; \mathbf{D} - \mathbf{A} = \mathbf{Q}'\Lambda' \mathbf{Q}'^{-1} \\
    \mathbf{GFT}_\mathbf{L} \coloneqq&\; \mathbf{Q}'^{-1} \\
    \label{eq:GFT_L}
    \mathbf{\tilde{x}} \coloneqq&\; \mathbf{GFT}_\mathbf{L} \mathbf{x}.
\end{align}

In the literature, the symmetrically normalised Laplacian $\mathbf{L}_\mathrm{norm}$ is sometimes used instead of the Laplacian \cite{ortega2018graph, gavili2017shift, sandryhaila2014discrete}. In the case of a graph $\mathbf{A}_c$ with linear topology, the degree matrix $\mathbf{D}_c$ is given by the identity matrix, which means that the symmetrically normalised Laplacian $\mathbf{L}_{c,\mathrm{norm}}$ of this graph is equivalent to the Laplacian $\mathbf{L}_c$:
\begin{align}
    \mathbf{L}_{c,\mathrm{norm}} = \mathbf{D}_c^{-1/2}\mathbf{L}_c\mathbf{D}_c^{-1/2}=\mathbbm{1}^{-1/2}\mathbf{L}_c\mathbbm{1}^{-1/2}=\mathbf{L}_c.
\end{align}
Consequently, the eigendecomposition of $\mathbf{L}_{c,\mathrm{norm}}$ is also given in terms of the DFT-matrix.

Lastly, note that eigenvectors $\mathbf{v}_k$ of $\mathbf{L}_c$ with $k>0$ are not the same as the eigenvectors of the symmetrised Laplacian $\mathbf{L}_{c,\mathrm{sym}}=\mathbf{D}_c - \left(\mathbf{A}_c + \mathbf{A}_c^\top\right)/2$: While the eigenvectors $\mathbf{v}_k$ are complex-valued for $k>0$, the eigenvectors of the $\mathbf{L}_{c,\mathrm{sym}}$ are necessarily real-valued due to the symmetry of $\mathbf{L}_{c,\mathrm{sym}}$. As a consequence, the eigendecomposition of $\mathbf{L}_{c,\mathrm{sym}}$ is not given in terms of the DFT-matrix; in other words, the eigendecomposition of the symmetrised Laplacian of a graph with linear topology does not reduce to the discrete Fourier transform.
However, given the analogy of $\mathbf{L}_{c,\mathrm{sym}}$ to the second partial time-derivative in the continuous case, of which the Fourier modes are also eigenfunctions, it may still be valid to use the symmetrised Laplacian $\mathbf{L}_{\mathrm{sym}}$ of an arbitrary graph structure for the GFT in lieu of $\mathbf{L}$. This may be especially useful if the measurement of the graph structure is intrinsically symmetric, while the actual graph structure is not. For example, when assessing the structural connectivity by measuring white matter between brain regions, the directionality cannot be determined due to limitations in the methodology, and only the symmetric graph is retrieved.

\indent

While equations (\ref{eq:GFT_A}) and (\ref{eq:GFT_L}) define the GFT for spatial signals, the GFT can be straightforwardly extended to multivariate signals $\mathbf{X}\in \mathbb{R}^{N_c\times N_t}$:
\begin{align}
    \tilde{\mathbf{X}} = \mathbf{GFT}_{\mathbf{A}/\mathbf{L}} \mathbf{X}.
\end{align}
In the matrix multiplication on the right-hand side, each column $\mathbf{x}_{*j}$, corresponding to a spatial signal at time $j$, is transformed successively.
The rows $\mathbf{\tilde{x}}_{i*}$ of the transformed multivariate signal $\mathbf{\tilde{X}}$ are the transformed signals, which are associated with the eigenvalues $\lambda_i$. Note that each transformed signal can also be thought of as a linear combination of the $N_c$ time signals $\mathbf{x}_{i*}$.

\subsection{Graph Fourier modes}
\label{ssec:graph_fourier_modes}
\onecolumngrid
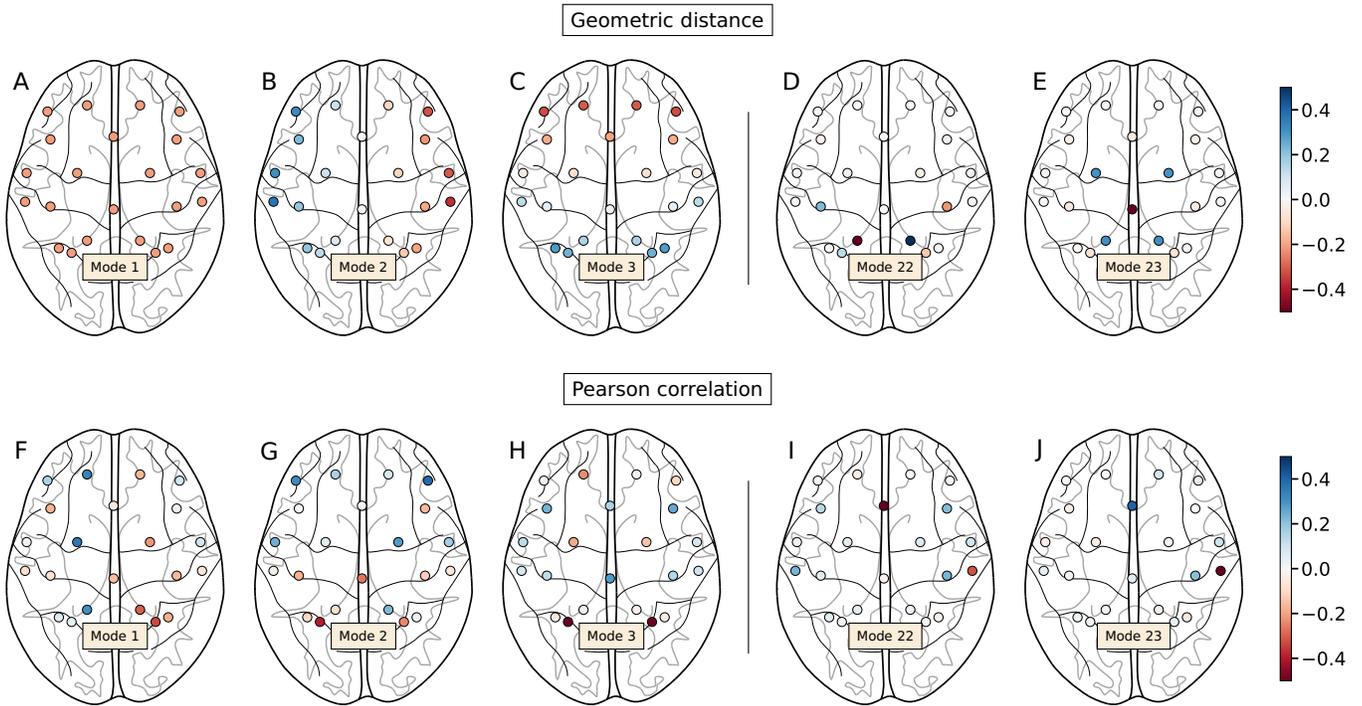
\begin{figure*}[t]
    \sffamily
    \begin{minipage}[c]{1\textwidth}
    \centering
    \begin{tikzpicture}
    \node[draw] at (0, 0) {\smaller Geometric distance};
    \end{tikzpicture}
    \vspace{5pt}
    \end{minipage}
    \noindent
    \begin{minipage}[c]{0.175\textwidth}
    \begin{tikzpicture}
    \draw (0, 0) node[inner sep=0] {\includegraphics[trim={3.2cm 0 3.2cm 0},clip,width=1\linewidth]{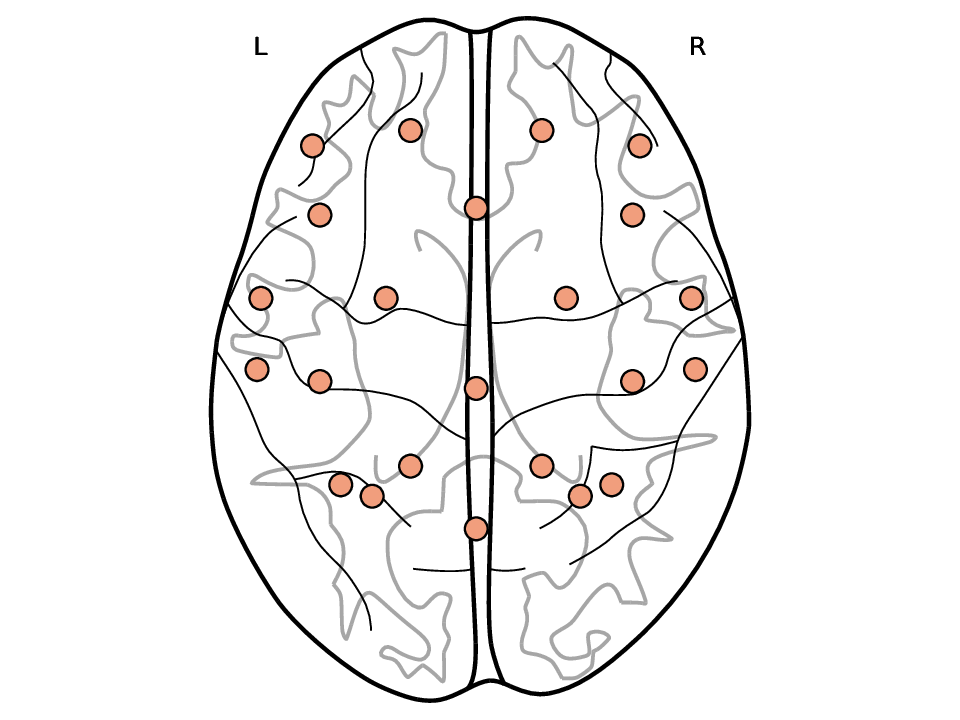}};
    \node[draw, fill=lightwheat] at (0,-0.92) {\tiny Mode 1};
    \node[fill=white] at (-1.265, 1.65) {\tiny \phantom{.}};
    \node[fill=white] at (+1.265, 1.65) {\tiny \phantom{.}};
    \draw (-1.25, 1.55) node {A};
    \end{tikzpicture}
    \end{minipage}\hfill
    \begin{minipage}[c]{0.175\textwidth}
    \begin{tikzpicture}
    \draw (0, 0) node[inner sep=0] {\includegraphics[trim={3.2cm 0 3.2cm 0},clip,width=1\linewidth]{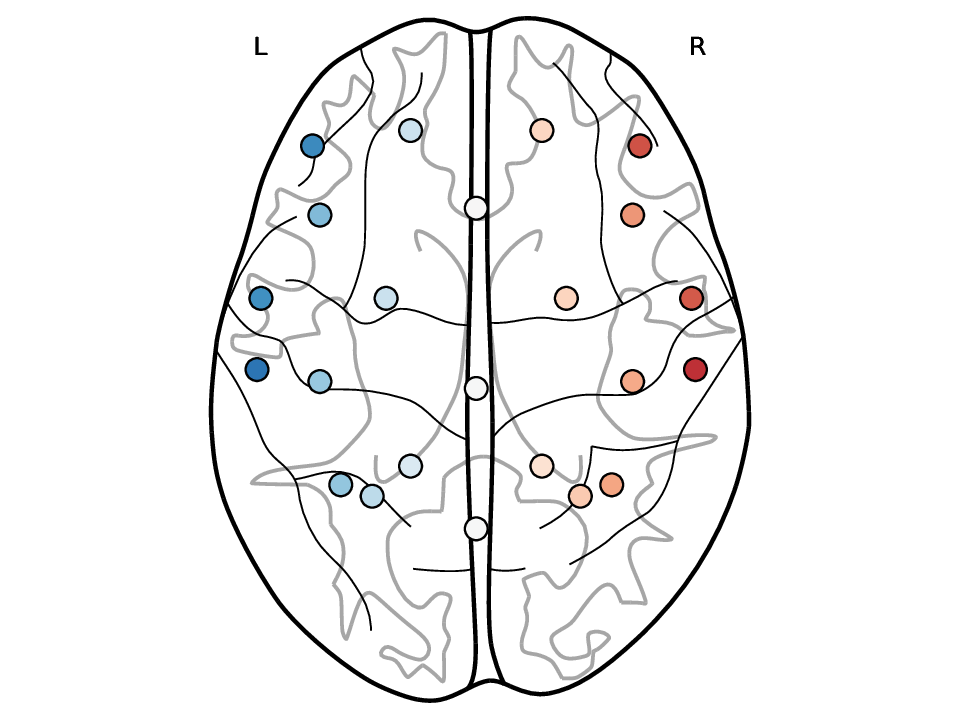}};     \node[draw, fill=lightwheat] at (0,-0.92) {\tiny Mode 2};
    \node[fill=white] at (-1.265, 1.65) {\tiny \phantom{.}};
    \node[fill=white] at (+1.265, 1.65) {\tiny \phantom{.}};
    \draw (-1.25, 1.55) node {B};
    \end{tikzpicture}
    \end{minipage}\hfill
    \begin{minipage}[c]{0.175\textwidth}
    \begin{tikzpicture}
    \draw (0, 0) node[inner sep=0] {\includegraphics[trim={3.2cm 0 3.2cm 0},clip,width=1\linewidth]{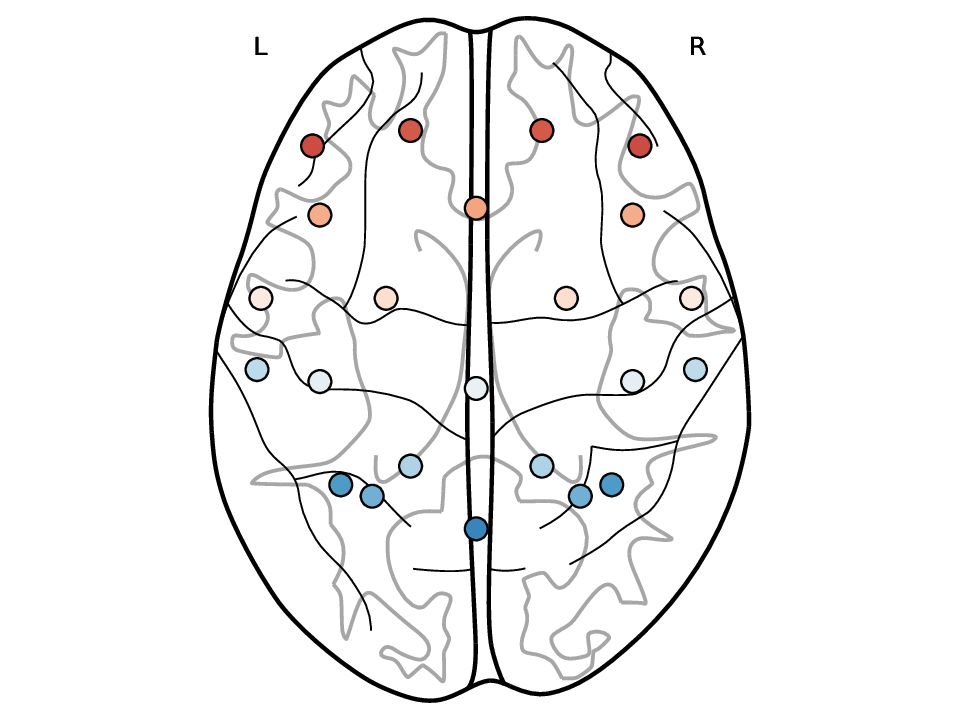}};     \node[draw, fill=lightwheat] at (0,-0.92) {\tiny Mode 3};
    \node[fill=white] at (-1.265, 1.65) {\tiny \phantom{.}};
    \node[fill=white] at (+1.265, 1.65) {\tiny \phantom{.}};
    \draw (-1.25, 1.55) node {C};
    \end{tikzpicture}
    \end{minipage}\hfill
    \begin{minipage}[t]{0.01\textwidth}
        \vspace{-35pt}
        \centering
        \tikz\draw[] (0,0) -- (0,-2.30);
    \end{minipage}\hfill
    \begin{minipage}[c]{0.175\textwidth}
    \begin{tikzpicture}
    \draw (0, 0) node[inner sep=0] {\includegraphics[trim={3.2cm 0 3.2cm 0},clip,width=1\linewidth]{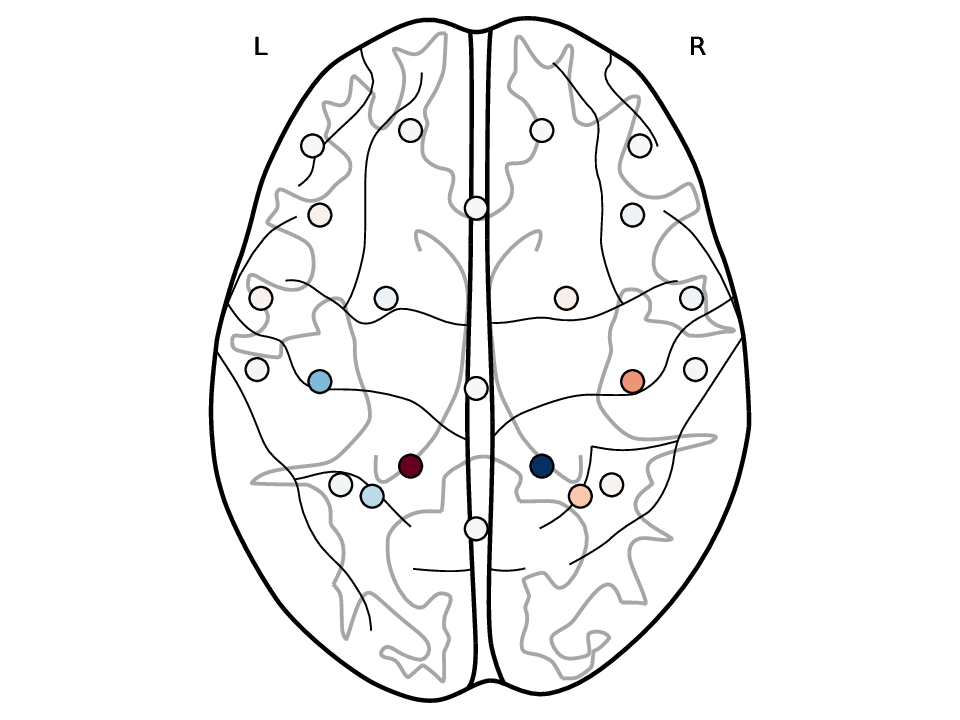}};     \node[draw, fill=lightwheat] at (0,-0.92) {\tiny Mode 22};
    \node[fill=white] at (-1.265, 1.65) {\tiny \phantom{.}};
    \node[fill=white] at (+1.265, 1.65) {\tiny \phantom{.}};
    \draw (-1.25, 1.55) node {D};
    \end{tikzpicture}
    \end{minipage}\hfill
    \begin{minipage}[c]{0.175\textwidth}
    \begin{tikzpicture}
    \draw (0, 0) node[inner sep=0] {\includegraphics[trim={3.2cm 0 3.2cm 0},clip,width=1\linewidth]{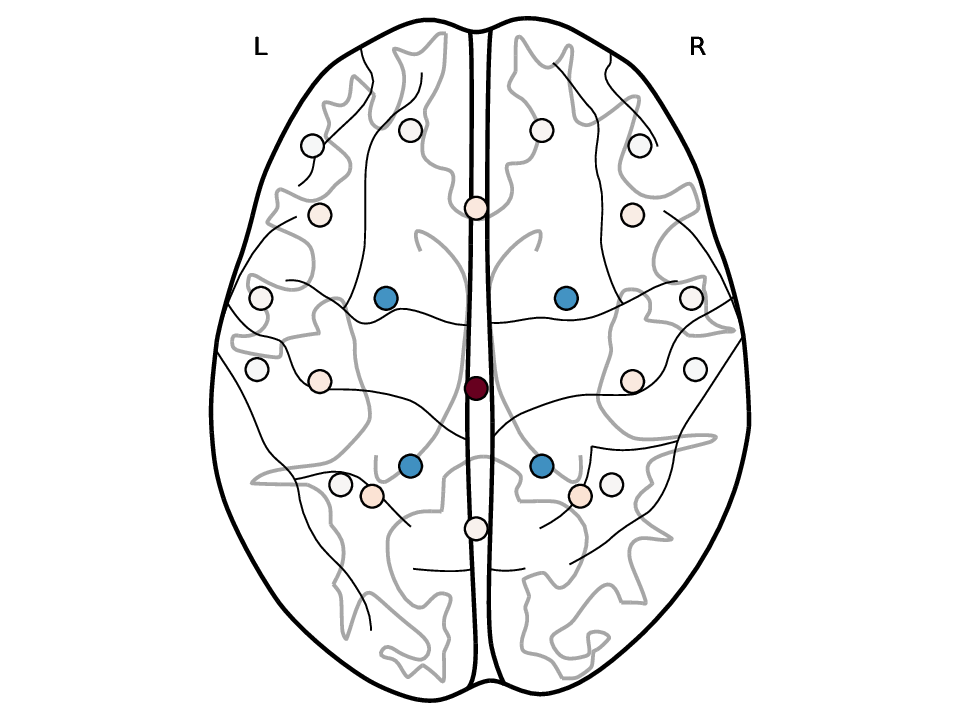}};     \node[draw, fill=lightwheat] at (0,-0.92) {\tiny Mode 23};
    \node[fill=white] at (-1.265, 1.65) {\tiny \phantom{.}};
    \node[fill=white] at (+1.265, 1.65) {\tiny \phantom{.}};
    \draw (-1.25, 1.55) node {E};
    \end{tikzpicture}
    \end{minipage}
    \hfill
    \begin{minipage}[c]{0.055\textwidth}
    \begin{tikzpicture}
    \draw (0, 0) node[inner sep=0] {\includegraphics[trim={12.4cm 0 0.8cm 0},clip,width=1\linewidth]{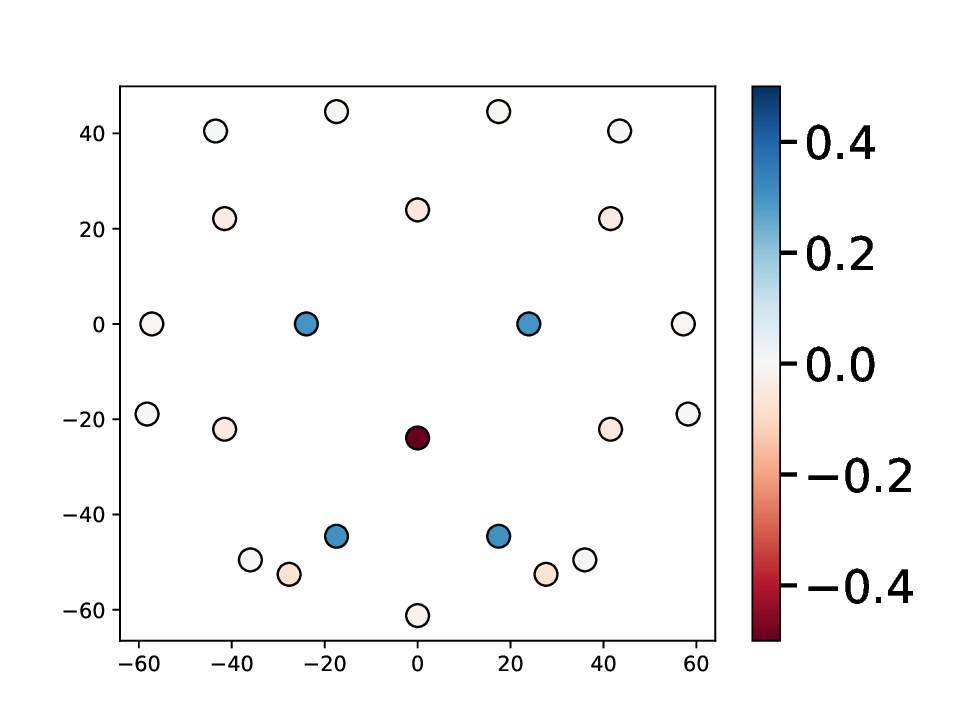}};
    \end{tikzpicture}
    \end{minipage}
    \begin{minipage}[c]
    {1\textwidth}
    \vspace{10pt}
    \centering
    \begin{tikzpicture}
    \node[draw] at (0, 0) {\smaller Pearson correlation};
    \end{tikzpicture}
    \vspace{5pt}
    \end{minipage}
    \noindent
    \begin{minipage}[c]{0.175\textwidth}
    \begin{tikzpicture}
    \draw (0, 0) node[inner sep=0] {\includegraphics[trim={3.2cm 0 3.2cm 0},clip,width=1\linewidth]{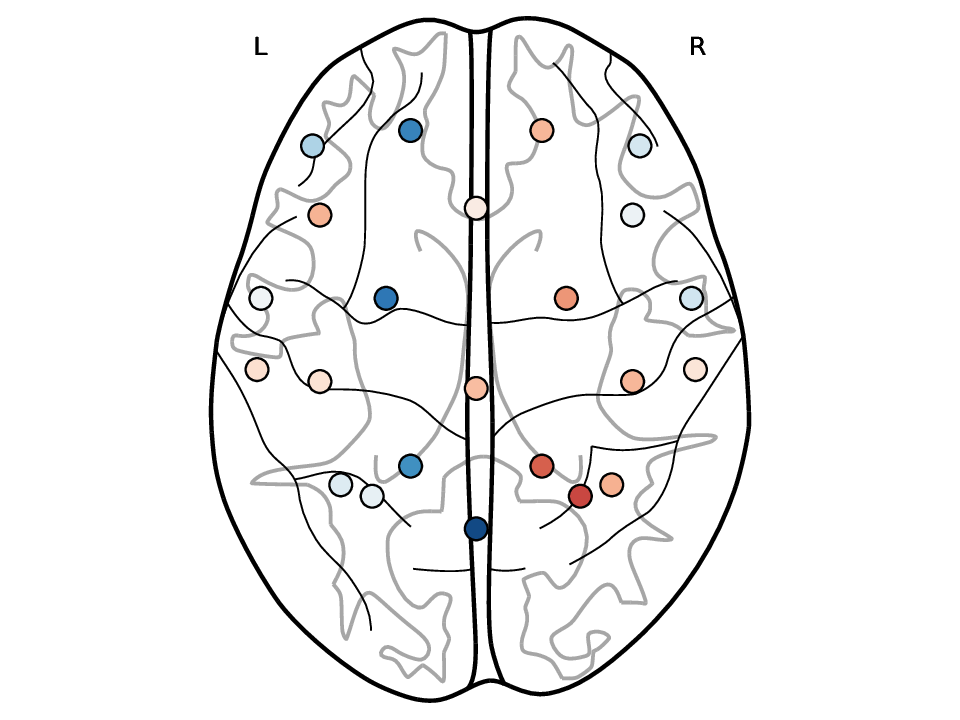}};     \node[draw, fill=lightwheat] at (0,-0.92) {\tiny Mode 1};
    \node[fill=white] at (-1.265, 1.65) {\tiny \phantom{.}};
    \node[fill=white] at (+1.265, 1.65) {\tiny \phantom{.}};
    \draw (-1.25, 1.55) node {F};
    \end{tikzpicture}
    \end{minipage}\hfill
    \begin{minipage}[c]{0.175\textwidth}
    \begin{tikzpicture}
    \draw (0, 0) node[inner sep=0] {\includegraphics[trim={3.2cm 0 3.2cm 0},clip,width=1\linewidth]{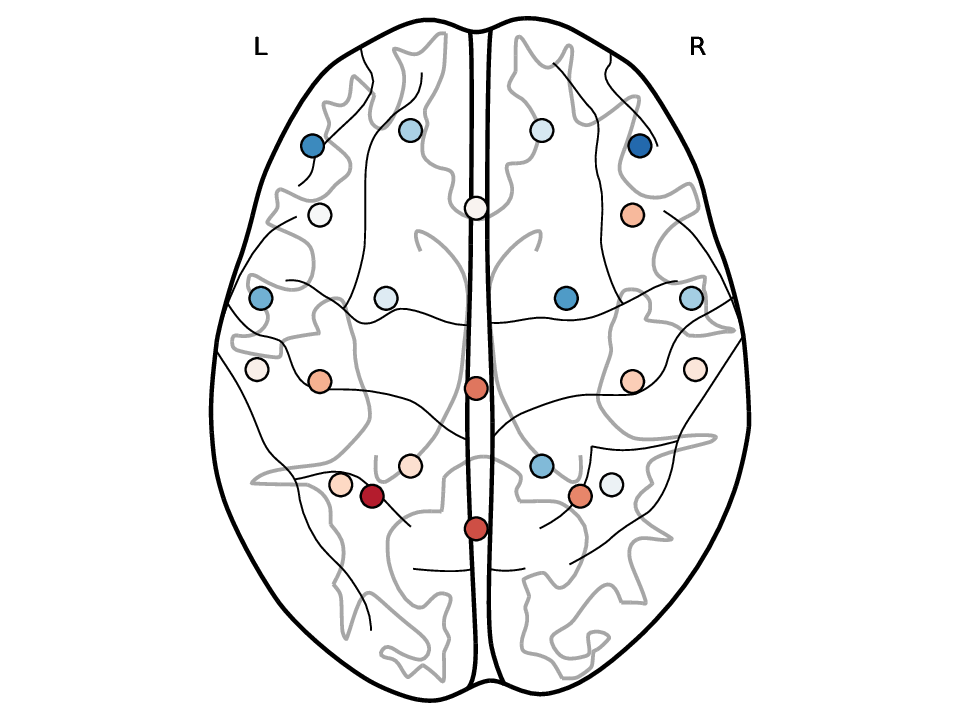}};     \node[draw, fill=lightwheat] at (0,-0.92) {\tiny Mode 2};
    \node[fill=white] at (-1.265, 1.65) {\tiny \phantom{.}};
    \node[fill=white] at (+1.265, 1.65) {\tiny \phantom{.}};
    \draw (-1.25, 1.55) node {G};
    \end{tikzpicture}
    \end{minipage}\hfill
    \begin{minipage}[c]{0.175\textwidth}
    \begin{tikzpicture}
    \draw (0, 0) node[inner sep=0] {\includegraphics[trim={3.2cm 0 3.2cm 0},clip,width=1\linewidth]{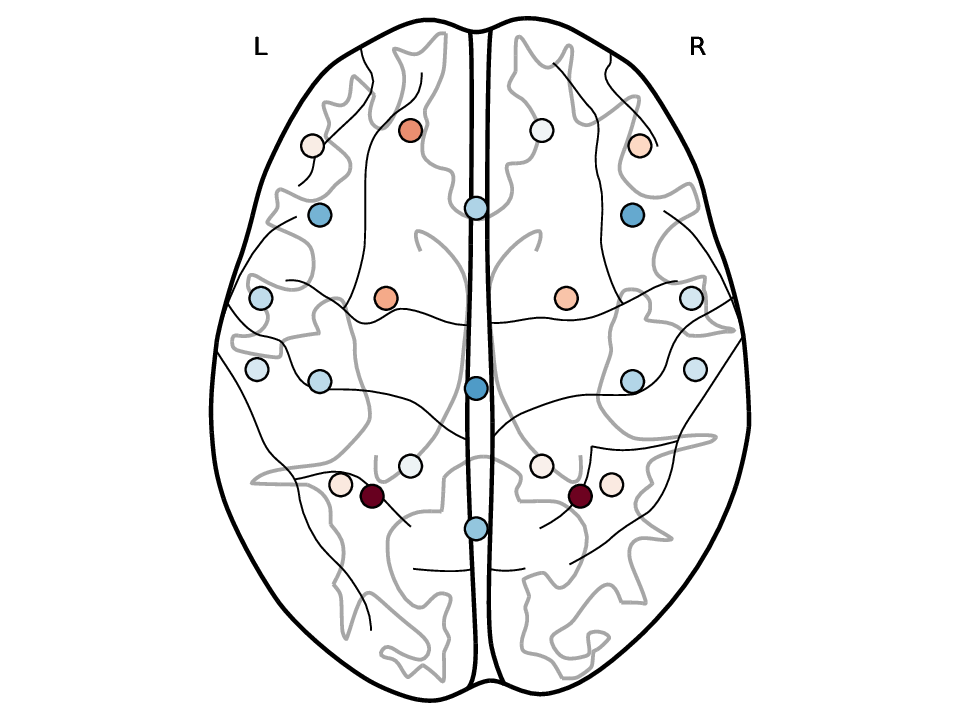}};     \node[draw, fill=lightwheat] at (0,-0.92) {\tiny Mode 3};
    \node[fill=white] at (-1.265, 1.65) {\tiny \phantom{.}};
    \node[fill=white] at (+1.265, 1.65) {\tiny \phantom{.}};
    \draw (-1.25, 1.55) node {H};
    \end{tikzpicture}
    \end{minipage}\hfill
    \begin{minipage}[t]{0.01\textwidth}
        \vspace{-35pt}
        \centering
        \tikz\draw[] (0,0) -- (0,-2.30);
    \end{minipage}\hfill
    \begin{minipage}[c]{0.175\textwidth}
    \begin{tikzpicture}
    \draw (0, 0) node[inner sep=0] {\includegraphics[trim={3.2cm 0 3.2cm 0},clip,width=1\linewidth]{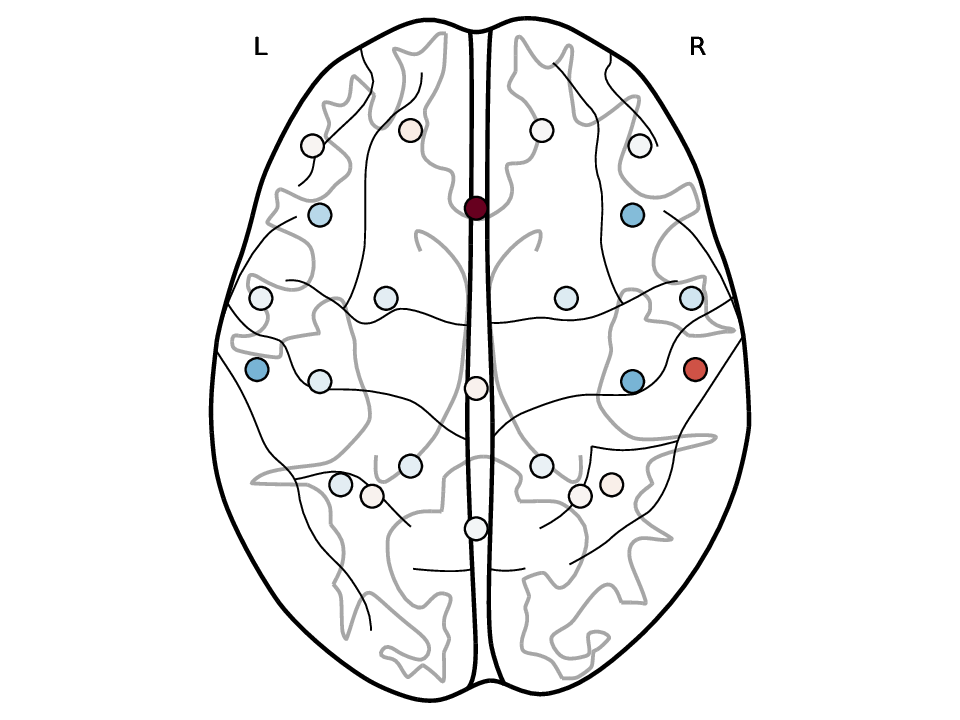}};     \node[draw, fill=lightwheat] at (0,-0.92) {\tiny Mode 22};
    \node[fill=white] at (-1.265, 1.65) {\tiny \phantom{.}};
    \node[fill=white] at (+1.265, 1.65) {\tiny \phantom{.}};
    \draw (-1.25, 1.55) node {I};
    \end{tikzpicture}
    \end{minipage}\hfill
    \begin{minipage}[c]{0.175\textwidth}
    \begin{tikzpicture}
    \draw (0, 0) node[inner sep=0] {\includegraphics[trim={3.2cm 0 3.2cm 0},clip,width=1\linewidth]{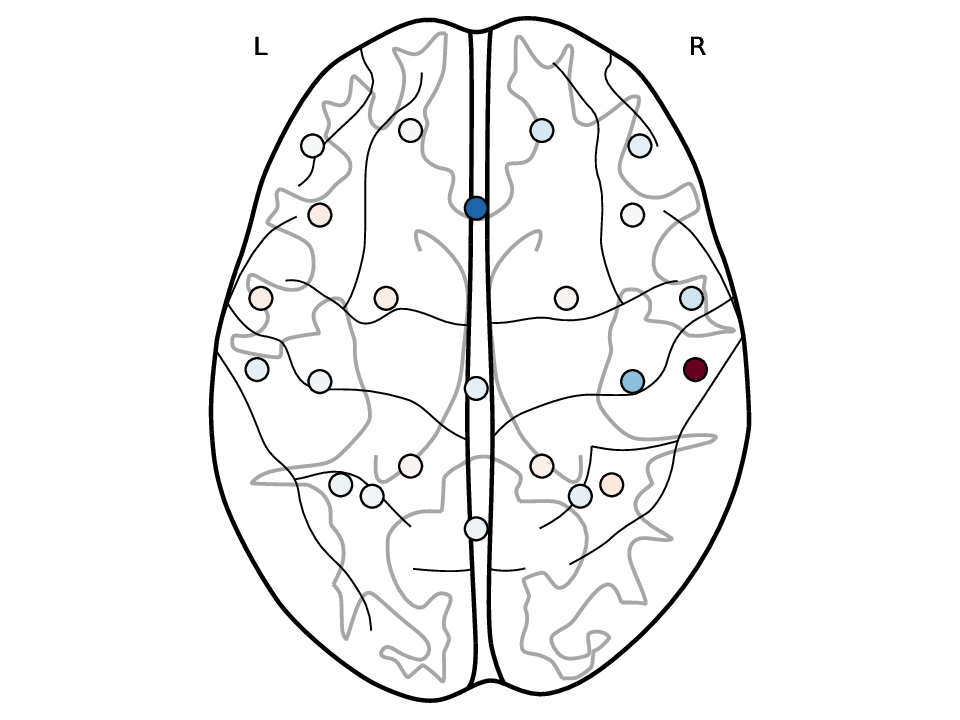}};     \node[draw, fill=lightwheat] at (0,-0.92) {\tiny Mode 23};
    \node[fill=white] at (-1.265, 1.65) {\tiny \phantom{.}};
    \node[fill=white] at (+1.265, 1.65) {\tiny \phantom{.}};
    \draw (-1.25, 1.55) node {J};
    \end{tikzpicture}
    \end{minipage}
    \hfill
    \begin{minipage}[c]{0.055\textwidth}
    \begin{tikzpicture}
    \draw (0, 0) node[inner sep=0] {\includegraphics[trim={12.4cm 0 0.8cm 0},clip,width=1\linewidth]{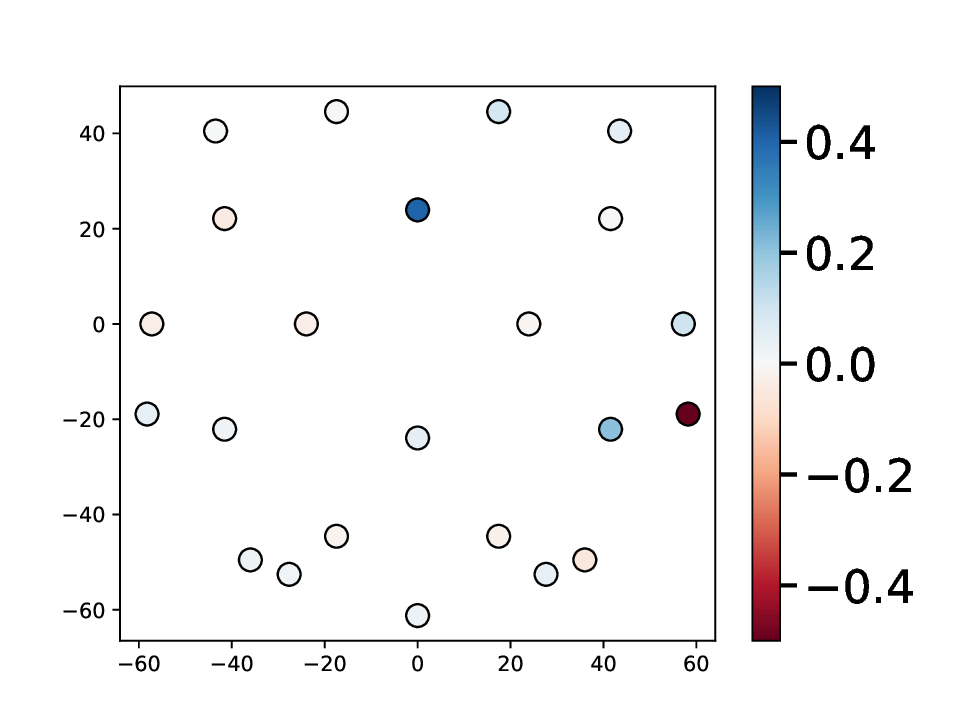}};
    \end{tikzpicture}
    \end{minipage}
    \caption{Lowest and highest graph Fourier modes for a geometric distance-based graph and a functional connectivity-based graph, computed from a real EEG data set. (A-C) The lowest graph Fourier modes of the geometric graph capture the fundamental symmetries and comprise the DC mode, the lateral symmetry mode and the coronal symmetry mode. (D-C) The highest graph Fourier modes are more localised. (F-H) The lowest graph Fourier modes of a functional connectivity Pearson correlation graph, which is computed from a real-world EEG data set. The graph contains negative weights, such that the DC mode is not necessarily the lowest mode. Modes 1 and 2 vaguely reflect the lateral symmetry mode 2 and the coronal symmetry mode 3 of the geometric  distance-based graph. (I-J) The highest graph Fourier modes of the Pearson correlation graph are localised and still exhibit a lateral symmetry}

\label{fig:graph_Fourier_modes}
\end{figure*}
\twocolumngrid
The GFT transforms a spatial signal $\mathbf{x}\in \mathbb{R}^{N_c}$ by projecting it onto the $N_c$ graph Fourier modes, which can be thought of as eigenmodes of the graph. This section intends to shed more light on the meaning of the graph Fourier modes.

Figure \ref{fig:graph_Fourier_modes} shows the graph Fourier modes for a real EEG data set. A description of the data set can be found in Klepl \textit{et al.} \cite{klepl2021characterising}. The upper row shows graph Fourier modes computed from the geometric distance between the EEG sensors, whereas the lower row shows modes computed from the functional connectivity, specifically the averaged pairwise Pearson correlation between the sensors.
Here, the relation between the graph Fourier modes and their corresponding frequency becomes more clear: Lower frequency modes appear as ``waves'' which are spread globally across the brain, whereas higher modes appear as highly localised and fast-varying waves. This general pattern is reproduced across the two graphs, even though they are retrieved by two fundamentally different methods.
One major difference between the graph Fourier modes of the two graphs is that the DC mode, or constant-value mode, is not the lowest frequency mode for the Pearson correlation graph. This is a consequence of the negative edge weights in the correlation-based graph, which induce eigenmodes with negative eigenvalues below the DC mode with eigenvalue zero.

The projections of a multivariate signal onto these eigenmodes, namely the GFT-trans\-formed signals, may yield insight into the acquired data beyond the raw signal. For example, the projection of the multivariate signal onto the DC mode corresponds to the mean of the signal across the channels.
The analysis of these transformed signals in terms of their usefulness for classification tasks is the main topic in the experimental part of this book chapter (see sections \ref{sec:methodology}-\ref{sec:discussion_conclusion}).

One major difference between Fourier modes in the DFT and graph Fourier modes in the GFT is that while the eigenvalues of the Fourier modes have a magnitude of one, the magnitudes of the eigenvalues of the graph Fourier modes are generally not equal to one.
When interpreting the GSO as a time evolution operator, it follows that the eigenmodes of the adjacency-based GFT are not stable in time, but either vanish or explode. This is in contrast to eigenmodes in Euclidean space, which are stable and energy-preserving. Examples from physics are electromagnetic waves or sound waves, of which the former can travel distances of billions of light years. 
Previously, an energy-preserving GSO was introduced by Gavili \textit{et al.} \cite{gavili2017shift}; however, this GSO is constructed by simply replacing the eigenvalues of the decomposed original GSO with eigenvalues of magnitude one. While this changes the original GSO, This leaves the GFT and thereby the eigenmodes untouched. While the new GSO preserves the energy of the eigenmodes, it may not represent the graph structure truthfully. 

\subsection{Total variation}
\label{ssec:total_variation}
A useful statistic of a graph signal $\mathbf{x}\in \mathbb{R}^{N_c}$ is its total variation (TV), which is a measure of how smooth the signal is across the graph structure $\mathbf{A}$.
The TV is based on the local variation, which in turn is based on the graph derivative of the graph signal. The graph derivative can be defined either on the node $i$ or on the edge $(i,j)$, leading to two different definitions of the TV, here referenced as the \textit{node-based} \cite{sandryhaila2014discrete} and the \textit{edge-based} \cite{shuman2013emerging} TV.

The node derivative calculates the difference between the signal and the signal shifted by the GSO, yielding a derivative at node $i$:
\begin{align}
    \nabla_i(\mathbf{x}) \coloneqq [\mathbf{x} - \mathbf{A}\mathbf{x}]_i = x_i - \sum_j a_{ij} x_j.
\end{align}
The local variation is then given by the magnitude $\Vert\nabla_i(\mathbf{x})\Vert_{l_1}$ of the graph derivative.

The edge derivative, on the other hand, weighs the difference between the signal at node $i$ and the signal at node $j$ by their connectivity, yielding a derivative along the edge $(i,j)$:
\begin{align}
    \nabla_{ij}(\mathbf{x}) \coloneqq \sqrt{a_{ij}}(x_i - x_j).
\end{align}
The local variation is then computed using the p-norm of the derivative vector $\nabla_{i*}(\mathbf{x})$:
\begin{align}
    \Vert\nabla_{i*}(\mathbf{x})\Vert_{l_p} = \left(\sum_{j} a_{ij}^{\frac{p}{2}}|x_i - x_j|^{p}\right)^{\frac{1}{p}},\quad p\in \mathbb{N}.
\end{align}

In both cases, the TV of a graph signal is computed from the sum of the local variation across all nodes. Given a graph structure $\mathbf{A}$, the node-based \cite{sandryhaila2014discrete} and the edge-based \cite{shuman2013emerging} TV of a graph signal $\mathbf{x}$ are then defined as follows:
\begin{align}
    \mathrm{TV}_\mathbf{A}^{(n)}(\mathbf{x}) &\coloneqq \sum_i \Vert\nabla_i(\mathbf{x})\Vert_{l_1} = \sum_i\Big|x_i - \sum_j a_{ij} x_j\Big|=\Vert \mathbf{x} - \mathbf{A}\mathbf{x}\Vert_1\\
    \label{eq:TV_edge}
    \mathrm{TV}_\mathbf{A}^{(e)}(\mathbf{x}) &\coloneqq \frac{1}{2}\sum_i \Vert\nabla_{i*}(\mathbf{x})\Vert_{l_2}^2 = \frac{1}{2}\sum_{i} \sum_{j} a_{ij}(x_i - x_j)^2.
\end{align}
The total variation for a multivariate signal $\mathbf{X}$, which comprises spatial signals for one time step each, is simply the sum of the total variation of all spatial signals in time:
\begin{align}
    \mathrm{TV}_\mathbf{A}^{(n)}(\mathbf{X}) &= \sum_{k=1}   ^{N_t}\sum_i\Big|x_{ik} - \sum_j a_{ij} x_{jk}\Big|=\Vert \mathbf{X} - \mathbf{A}\mathbf{X}\Vert_{1,1}\\
    \mathrm{TV}_\mathbf{A}^{(e)}(\mathbf{X}) &= \sum_{k=1}   ^{N_t}\frac{1}{2} \sum_{i}\sum_{j} a_{ij}(x_{ik} - x_{jk})^2,
\end{align}
where $\Vert \cdot \Vert_{1,1}$ denotes the entry-wise $L_{1,1}$ matrix norm.

Note that the node-based TV can also be computed from the normalised adjacency matrix $\mathbf{A}_\mathrm{norm}=\mathbf{A} / |\lambda_\mathrm{max}|$, where $\lambda_\mathrm{max}$ is the largest eigenvalue of $\mathbf{A}$, which ensures numerical stability. While the node-based TV is always either positive or zero, the edge-based TV can become negative if we allow negative weights $a_{ij}<0$ in the adjacency matrix.
Section \ref{ssec:graph_retrieval} discusses the validity of negative weights and possible implications.

The TV allows to understand the ordering of the graph eigenvectors as frequencies in terms of their eigenvalue. In the case of the node-based TV with $\mathbf{A}_\mathrm{norm}$, the TV for a normalised eigenvector $\mathbf{v}_k$ with real eigenvalue $\lambda_k$ is given by:
\begin{align}
    \mathrm{TV}_{\mathbf{A}_\mathrm{norm}}^{(n)}(\mathbf{v}_k) &= \Big|1 - \frac{\lambda_k}{\lambda_\mathrm{max}}\Big|. 
\end{align}
It can easily be seen that the following holds for two eigenvectors $\mathbf{v}_k$ and $\mathbf{v}_l$ of $\mathbf{A}_\mathrm{norm}$ with respective real eigenvalues $\lambda_k$ and $\lambda_l$:
\begin{align}
    \lambda_k &< \lambda_l\\ \Rightarrow \mathrm{TV}_{\mathbf{A}_\mathrm{norm}}^{(n)}(\mathbf{v}_k) &> \mathrm{TV}_{\mathbf{A}_\mathrm{norm}}^{(n)}(\mathbf{v}_l).
\end{align}
In other words, the ordering of the eigenvalues from highest to lowest orders the graph eigenvectors from lowest to highest frequency. Notice that any eigenvector $\mathbf{v}_k$ of $\mathbf{A}_\mathbf{norm}$ is also an eigenvector of $\mathbf{A}$.

A similar relationship between eigenvalue and frequency can be established for the edge-based TV if we assume the adjacency matrix to be symmetric, i.d. $a_{ij}=a_{ji}$. To this end, the TV is firstly expressed in terms of the Laplacian:
\begin{align}
    \mathrm{TV}_\mathbf{A}^{(e)}(\mathbf{x}) &= \frac{1}{2}\sum_{i,j} a_{ij}(x_i - x_j)^2 \\&= \frac{1}{2}\sum_{i,j} a_{ij}x_i^2 + \frac{1}{2}\sum_{i,j} a_{ij}x_j^2 - \frac{1}{2}\sum_{i,j} 2 a_{ij}x_ix_j\\
    &\stackrel{\mathmakebox[\widthof{=}]{a_{ij}=a_{ji}}}{=}\ \ \ \sum_{i,j} a_{ij}x_i^2 - \sum_{i,j} a_{ij}x_ix_j\\
    &= \mathbf{x}^\top 
  \begin{bmatrix}
    \sum_j a_{1j} & & \\
    & \ddots & \\
    & & \sum_j a_{N_cj}
  \end{bmatrix}\mathbf{x} - \mathbf{x}^\top \mathbf{A} \, \mathbf{x}\\
  &=\mathbf{x}^\top (\mathbf{D} - \mathbf{A}) \,\mathbf{x}=\mathbf{x}^\top \mathbf{L} \,\mathbf{x}.
\end{align}

This representation of the TV allows to link the eigenvalues to graph frequencies. For a normalised eigenvector $\mathbf{v}_k$ of $\mathbf{L}$ with eigenvalue $\lambda_k$, the TV is given by:
\begin{align}
    \mathrm{TV}_\mathbf{A}^{(e)}(\mathbf{v}_k) = \mathbf{v}_k^\top \mathbf{L} \mathbf{v}_k = \lambda_k \Vert \mathbf{v}_k \Vert_2 = \lambda_k.
\end{align}
Hence, the frequency of the eigenvector $\mathbf{v}_k$ of $\mathbf{L}$, which is given in terms of its TV, is directly linked to its eigenvalue. Importantly, here the eigenvectors of the Laplacian matrix act as the graph frequencies, while in the case of the node-based TV the eigenvectors of the adjacency matrix were taken to be the graph frequencies. This explains why here higher eigenvalues are linked to higher graph frequencies, while in the case of the node-based TV the relationship is inverse.

Using the eigendecomposition of the Laplacian matrix, $\mathbf{L} = \mathbf{GFT}^\top \Lambda \mathbf{GFT}$, the expression can be further rewritten as follows:
\begin{align}
    \mathrm{TV}_\mathbf{A}^{(e)}(\mathbf{x})&=\mathbf{x}^\top \mathbf{L} \,\mathbf{x} = \mathbf{x}^\top \mathbf{GFT}^\top \Lambda \mathbf{GFT} \,\mathbf{x}\\
    &= \tilde{\mathbf{x}}^\top\Lambda\,\tilde{\mathbf{x}} = \sum_k \lambda_k \tilde{x}_k^2.
\end{align}
This identity of the edge-based TV allows to understand how it can be decomposed in terms of their graph frequency components $\tilde{x}_k$.

\subsection{Graph convolution}
\label{ssec:graph_convolution}
The definition of the graph convolution is built on the ...
In the classical case, the convolution for two period time signals $s$ and $r$, represented by the time-ordered vectors $\mathbf{s}=(s_0,...,s_{N-1})^\top$ and $\mathbf{r}=(r_0,...,r_{N-1})^\top$, is given by the vector
\begin{align}
    \Big((s*r)[n]\Big)_{1\leq n\leq N}^\top = \left(\sum_i r_i \mathbf{A}_c^i\right) \mathbf{s}.
\end{align}
The following two identities can be directly derived from the eigendecomposition of $\mathbf{A}_c$ in equations (\ref{eq:Ac_eigendecomposition}) and (\ref{eq:Ac_DFT}):
\begin{align}
    \mathbf{A}_c^i &= \mathbf{DFT}^{-1}\Lambda_c^i\, \mathbf{DFT}\\
    \label{eq:Lambda_c_DFT}
    \sum_i r_i \Lambda_c^i &= \mathrm{diag}(\mathbf{DFT}\,\mathbf{r}/\sqrt{N}).
\end{align}
Using these two identities, the vector of the convoluted signal $s*r$ can be rewritten as follows:
\begin{align}
    \Big((s*r)[n]\Big)_{1\leq n\leq N}^\top &= \sum_i r_i\, \mathbf{DFT}^{-1} \Lambda_c^i\,\mathbf{DFT}\,\mathbf{s}\\
    &= \mathbf{DFT}^{-1} \left(\sum_i r_i\, \Lambda_c^i\right)\,\mathbf{DFT}\,\mathbf{s}\\
    &= \mathbf{DFT}^{-1} \mathrm{diag}\left(\mathbf{DFT}\,\mathbf{r}/\sqrt{N}\right)\,\mathbf{DFT}\,\mathbf{s}\\
    &= \mathbf{DFT}^{-1} \frac{1}{\sqrt{N}}(\mathbf{DFT}\,\mathbf{r})\circ\,(\mathbf{DFT}\,\mathbf{s}).
\end{align}

The analogy between the classical shift operator and the GSO can be used to define the graph convolution on two spatial signals $\mathbf{x}$ and $\mathbf{y}$:
\begin{align}
    \label{eq:convolution_A}
    \mathbf{x}*\mathbf{y} = \left(\sum_i y_i \mathbf{A}^i\right) \mathbf{x}.
\end{align}
Using the filter-based GFT, we can use the identity
\begin{align}
\label{eq:A_exp_identity}
    \mathbf{A}^i = \mathbf{GFT}_\mathbf{A}^{-1}\Lambda^i\,\mathbf{GFT}_\mathbf{A},
\end{align}
which can be trivially derived from (\ref{eq:A_GFT}), to rewrite the graph convolution:
\begin{align}
    \mathbf{x}*\mathbf{y} &= \sum_i y_i\,\mathbf{GFT}_\mathbf{A}^{-1} \Lambda^i\,\mathbf{GFT}_\mathbf{A}\,\mathbf{x}\\
    &= \mathbf{GFT}_\mathbf{A}^{-1}\left(\sum_i y_i\Lambda^i\right)\mathbf{GFT}_\mathbf{A}\,\mathbf{x}\\
    &= \mathbf{GFT}_\mathbf{A}^{-1}\mathrm{diag}\left(\sum_i y_i\lambda_0^i,...,\sum_i y_i\lambda_{N-1}^i\right)\,\mathbf{GFT}_\mathbf{A}\,\mathbf{x}.
\end{align}
The crucial difference between the classical convolution and the graph convolution is that the identity (\ref{eq:Lambda_c_DFT}) does not have a comparable equivalent in GSP.

We can also implicitly define the graph convolution for the case of the derivative-based GFT by analogy to the classical convolution:
\begin{align}
    \mathbf{x}*\mathbf{y} &= \mathbf{GFT}_\mathbf{L}^{-1}\left(\sum_i y_i\Lambda'^{\,i}\right)\mathbf{GFT}_\mathbf{L}\,\mathbf{x}\\
    \label{eq:convolution_L}
    &=\left(\sum_i y_i \mathbf{L}^i\right) \mathbf{x}.
\end{align}
This definition of the graph convolution has been commonly used in the literature \cite{defferrard2016convolutional,kipf2016semi,li2018deeper,zhang2019graph}.
However, comparing the last equation (\ref{eq:convolution_L}) to the GSO-based convolution in equation (\ref{eq:convolution_A}), we can infer that this definition implicitly assumes the Laplacian matrix to be a GSO, which may not be an adequate use of the Laplacian matrix.

\subsection{Graph signal filtering}
\label{ssec:graph_signal_filtering}
In classical signal processing, filters can be described both by their impulse response in the time domain as well as by their frequency response in the Fourier frequency domain; the two descriptions are linked by the DFT. Graph filters can be similarly defined by their impulse and by their frequency response; here the descriptions are linked by the GFT. The former, impulse response description was previously encountered in subsection \ref{sssec:filter_GFT} and is closely related to the concept of convolution as introduced in the previous subsection \ref{ssec:graph_convolution}. The frequency response description, however, allows to define graph spectral band-pass filters, which is essential for tasks such as graph denoising. For example, a low-pass filter can be used to reduce the total variation of the spatial signal.

In the time domain, a filter $\mathbf{H}$ can be built by accessing previous signal values through shifting the signal $\mathbf{x}$ and weighing each of these values by $p_i$, as previously shown in equation (\ref{eq:H_through_time}):
\begin{align}
\label{eq:H_through_time_classical}
    \mathbf{H}\mathbf{x} = \left(\sum_{i=0}^{N-1}p_i \mathbf{A}_c^i\right)\mathbf{x}= h(\mathbf{A}_c) \mathbf{x}=\tilde{\mathbf{x}},
\end{align}
where $\tilde{\mathbf{x}}$ is the filtered graph signal and $h$ is the polynomial filter function. 

Using the identity (\ref{eq:A_exp_identity}) with $\mathbf{A}=\mathbf{A}_c$,
\begin{align}
    \mathbf{A}_c^i = \mathbf{DFT}^{-1}\Lambda_c^i\,\mathbf{DFT},
\end{align}
we can rewrite equation (\ref{eq:H_through_time_classical}):
\begin{align}
    \mathbf{H}\mathbf{x} &= \mathbf{DFT}^{-1} \left(\sum_{i=0}^{N-1}p_i \Lambda_c^i\right)\mathbf{DFT}\,\mathbf{x}=\mathbf{DFT}^{-1}h(\Lambda_c)\,\mathbf{DFT}\,\mathbf{x}\\
    &\eqqcolon \mathbf{DFT}^{-1}\mathrm{diag}\left(h_0, ..., h_{N-1}\right)\,\mathbf{DFT}\,\mathbf{x}.
\end{align}
In other words, to define the filter $\mathbf{H}$ we can either use the $N$ parameters $p_i$, which define the filter by its weights in the time domain, or equivalenty the parameters $h_i$, which define it by its weights in the Fourier domain.

In analogy to this classical case, we can define a graph filter $\mathbf{H}_{\mathbf{A}/\mathbf{L}}$ in the spatial domain for the two definitions of the GFT as follows:
\begin{align}
    \mathbf{H}_\mathbf{A} \mathbf{x}&= \left(\sum_{i=0}^{N-1}p_i \mathbf{A}^i\right)\mathbf{x}\\
    \mathbf{H}_\mathbf{L} \mathbf{x}&= \left(\sum_{i=0}^{N-1}p_i \mathbf{L}^i\right)\mathbf{x}.
\end{align}
Note that the Laplacian-based GFT filter is not built on the graph convolution, whereas the adjacency matrix-based GFT filter is.

As in the classical case, the same filters can be described in the graph spectral domain using the eigenvalue matrix $\Lambda_{\mathbf{A}/\mathbf{L}}$:
\begin{align}
    \mathbf{H}_\mathbf{A} \mathbf{x}&= \mathbf{GFT}_\mathbf{A}^{-1} h(\Lambda_\mathbf{A}) \mathbf{GFT}_\mathbf{A} \mathbf{x}\\&\eqqcolon \mathbf{GFT}_\mathbf{A}^{-1} \mathrm{diag}\left(h_0^{(A)},...,h_{N-1}^{(A)}\right) \mathbf{GFT}_\mathbf{A} \mathbf{x}\\
    \mathbf{H}_\mathbf{L} \mathbf{x}&= \mathbf{GFT}_\mathbf{L}^{-1} h(\Lambda_\mathbf{L}) \mathbf{GFT}_\mathbf{L} \mathbf{x}\\&\eqqcolon \mathbf{GFT}_\mathbf{L}^{-1} \mathrm{diag}\left(h_0^{(L)},...,h_{N-1}^{(L)}\right) \mathbf{GFT}_\mathbf{L} \mathbf{x}.
\end{align}

The definition of the filter $\mathbf{H}$ can be extended to multivariate signals $\mathbf{X}\in \mathbb{R}^{N_c\times N_t}$ as follows:
\begin{align}
    \mathbf{H}_{\mathbf{A/\mathbf{L}}}\, \mathbf{X} = \mathbf{X}_F.
\end{align}
The parameters $p_i$ or $h_i$ can either be designed or learned using machine learning-approaches. Sometimes, the spatial domain description of the filter is used to avoid computing the eigendecomposition of the adjacency matrix or the Laplacian xxx cite. To further limit computations or the number of parameters, only the first $k$ parameters $p_i$ are used, such that $p_i=0$ for $i>=k$.

\subsection{Spectral graph wavelets}
\label{ssec:spectral_graph_wavelets}
The concept of wavelet transforms can be transferred to graphs as well. In classical signal processing, wavelets for discrete periodic signals $\mathbf{s}$ are scalable and time-localised vectors $\psi_{s,\tau}\in \mathbb{R}^N$, where $0<s<1$ indicates the scaling factor and $\tau\in \{1,...,N_t\}$ indicates at which time point the wavelet is located. The wavelets can be defined either with bidirectional or unidirectional scaling:
\begin{align}
    \psi_{s, \tau}^{(\mathrm{bi})} &\coloneqq \left(\sum_{k=0}^{N-1}p_k s^{N/2-|k-N/2|}\mathbf{A}_c^k\right)\delta_\tau=\left(\sum_{k=0}^{N-1}\tilde{p}_k\mathbf{A}_c^k\right)\delta_\tau\\
    \psi_{s, \tau}^{(\mathrm{uni})} &\coloneqq \left(\sum_{k=0}^{N-1}p_k(s\mathbf{A}_c)^k\right)\delta_\tau=\left(\sum_{k=0}^{N-1}\hat{p}_k\mathbf{A}_c^k\right)\delta_\nu=h(s\mathbf{A}_c)\delta_\tau.
\end{align}
Here, the delta impulse $\delta_\tau\in \mathbb{R}^N_t$, which is $\left( \delta_\tau \right)_\kappa=1$ at $\kappa=\tau$ and $\left( \delta_\tau \right)_\kappa=0$ elsewhere, localises the wavelet at time step $\tau$. Wavelet values $k$-steps away from $\tau$ are suppressed by a factor $s^{N/2-|k-N/2|}$ or $s^k$ for bi- or unidirectional wavelets, respectively. This means that values far away from time $\tau$ are suppressed more strongly, whereby the distance is measured either to either side of $\tau$ (bidirectional), or backwards in time (unidirectional).

In analogy to classical discrete wavelets, spectral graph wavelets can be constructed using the concepts of graph convolution and graph signal filtering. 
Previous definitions are built on the Laplacian-based graph filtering and solely applied unidirectional scaling \cite{hammond2011wavelets,tremblay2014graph,dong2020graph}. The graph spectral wavelets for node $\nu$, built on the adjacency matrix- and Laplacian-based graph filtering with unidirectional scaling, are given by:
\begin{align}
    \psi_{s, \nu}^{(A)} &\coloneqq \left(\sum_{k=0}^{N-1}p_k(s\mathbf{A})^k\right)\delta_\nu=h(s\mathbf{A})\delta_\nu\\
    \psi_{s, \nu}^{(L)} &\coloneqq \left(\sum_{k=0}^{N-1}p_k(s\mathbf{L})^k\right)\delta_\nu=h(s\mathbf{L})\delta_\nu.
\end{align}
The scaling factor $s$ can be factored into the impulse response values $p_k$, yielding the following expressions:
\begin{align}
    \psi_{s, \nu}^{(A)} &=\left(\sum_{k=0}^{N-1}\tilde{p}_k\mathbf{A}^k\right)\delta_\nu=\tilde{h}(\mathbf{A})\delta_\nu=\mathbf{\tilde{H}}_\mathbf{A} \delta_\nu\\
    \psi_{s, \nu}^{(L)} &=\left(\sum_{k=0}^{N-1}\tilde{p}_k\mathbf{L}^k\right)\delta_\nu=\tilde{h}(\mathbf{L})\delta_\nu=\mathbf{\tilde{H}}_\mathbf{L} \delta_\nu.
\end{align}
From these expressions, it is clear that the spectral graph wavelets are given as the columns of a filter with matrix $\mathbf{\tilde{H}}_{\mathbf{A}/\mathbf{L}}$, which is defined by its impulse response values $\tilde{p}_k$.

\subsection{Links between graph retrieval methods}
\label{ssec:link_TV_FC}
Given the plethora of graph retrieval methods, the question arises as to under what conditions some of the graph retrieval methods are equivalent. 
This section specifically looks at the links between functional connectivity-based graphs and total variation-optimised graphs.

Kalofolias already showed such a link for a commonly used graph $\mathbf{A}_\mathrm{exp}\in \mathbb{R}^{N\times N}$, which is constructed from the rows $\mathbf{x}_{i*}$ of the data matrix $\mathbf{X}$ as follows \cite{kalofolias2016learn}:
\begin{align}
    a_{ij}^{(\mathrm{exp})} = \exp\left(-\frac{\Vert \mathbf{x}_{i*} - \mathbf{x}_{j*}\Vert_2^2}{\sigma}\right).
\end{align}
Given the following regularisation term,
\begin{align}
    f^{(\mathrm{log})}(\mathbf{A}) = \sigma^2 \sum_i\sum_j a_{ij}\left(\log\left(a_{ij}\right)-1\right),
\end{align}
this functional connectivity-based connectivity matrix minimises the edge-based total variation:
\begin{align}
    \underset{\mathbf{A}\in \mathcal{A}}{\arg\!\min}\,\,\mathrm{TV}_\mathbf{A}^{(e)}(\mathbf{X}) + f^{(\mathrm{log})}(\mathbf{A})=\mathbf{A}_\mathrm{exp},
\end{align}
where $\mathcal{A}$ denotes the set of real-valued $N \times N$ adjacency matrices.

In the following, suppose we are given a normalised multivariate signal $\mathbf{X}_\mathrm{norm}$, where each row $\mathbf{x}_{i*}$, corresponding to the time signal at sensor $i$, is standardised to mean zero and standard deviation one.
We show that for such a signal a similar link between a total variation-optimised graph and a functional connectivity-based graph, whose entries in the adjacency matrix are the pairwise Pearson correlations, can be established. Note that due to the normalisation of the time signals, this Pearson correlation matrix is equivalent to the covariance matrix.

We firstly introduce the regularisation term
\begin{align}
\label{eq:Frobenius_reg}
    f^{(\mathrm{c})}(\mathbf{A}) = \frac{1}{2}\left\Vert \mathbf{J}_N - \mathbf{A}\right\Vert _F^2=\frac{1}{2}\sum_{i}\sum_{j}(1-a_{ij})^2,
\end{align}
where $\mathbf{J}_N$ is an all-ones matrix of size $N \times N$, and $\Vert \cdot \Vert_F$ denotes the Frobenius norm. Specifically, this regularisation term aims to move the entries $a_{ij}$ of $\mathbf{A}$ closer to 1.

Then, the solution to the optimisation problem
 \begin{align}
 \label{eq:optimisation_problem}
    \underset{\mathbf{A}\in \mathcal{A}}{\arg\!\min}\,\,\mathrm{TV}_\mathbf{A}^{(e)}(\mathbf{X}_\mathrm{norm}) + f^{(\mathrm{c})}(\mathbf{A})
\end{align}
 is the adjacency matrix given by the correlation matrix $\mathbf{A}_c$, where the entries $a_{ij}^{(c)}$ are calculated as the Pearson correlation between the normalised time signals $\mathbf{x}_{i*}$ and $\mathbf{x}_{j*}$ of the sensors $i$ and $j$, respectively. 
 This can be shown by taking the derivative with respect to $a_{ij}$:
 \begin{align}
     \frac{\partial}{\partial a_{ij}}&\mathrm{TV}_\mathbf{A}^{(e)}(\mathbf{X}_\mathrm{norm}) + f^{(\mathrm{c})}(\mathbf{A})\Big|_{\mathbf{A}=\mathbf{A}^\mathrm{max}}\\=&\frac{\partial}{\partial a_{ij}} \sum_{k=1}^{N_t}\sum_{i,j}\left( a_{ij}x_{ik}^2 -  a_{ij}x_{ik}x_{jk} + \frac{1}{2}(1-a_{ij})^2\right)\Big|_{a_{ij}=a_{ij}^\mathrm{max}}\\
     =&\sum_{k=1}   ^{N_t}x_{ik}^2-x_{ik}x_{jk}+(1-a_{ij}^\mathrm{max})(-1)\\
     =&\sum_{k=1}^{N_t}1-x_{ik}x_{jk}-1+a_{ij}^\mathrm{max}\\
     =&\left(-\sum_{k=1}^{N_t}x_{ik}x_{jk}\right)+N_t a_{ij}^\mathrm{max}=0\\
     &\Rightarrow \,a_{ij}^\mathrm{max} =\frac{1}{N_t}\sum_{k=1}^{N_t}x_{ik}x_{jk}=a_{ij}^{(c)}.
 \end{align}
 
In other words, the correlation matrix $\mathbf{A}^{(c)}$ minimises the total edge-based variation of a normalised data matrix $\mathbf{X}$ given the regularisation term defined in (\ref{eq:Frobenius_reg}). Therefore, it can be equally retrieved by learning the solution to the optimisation problem in (\ref{eq:optimisation_problem}).
Note that the requirement that the signals have to be normalised is naturally given for many neurophysiological signals, such as EEG signals. In our experiment, starting from section \ref{sec:methodology}, we normalise the data and use the Pearson correlation matrix as the graph for the data.

\subsection{Links between graph representations}
\label{ssec:rel_adj_lap}
There are arguments for both using the adjacency or the Laplacian matrix as the GSO in GSP, and both are frequently utilised in the literature.
In a direct comparison, Huang \textit{et al.} have observed no noticeable difference between utilising the adjacency or the Laplacian matrix as the GSO \cite{huang2018graph_perspective}. This observation poses the question of the relation between the use of the adjacency and the Laplacian matrix in GSP.
We demonstrate that for large graphs with normally distributed adjacency matrix weights, the two GSOs have similar eigenvectors. This in turn would mean that their respective GFTs are similar.

We here assume that the weights of the adjacency matrix follow a normal distribution with mean $\mu$ and standard deviation $\sigma$. 
Accordingly, the entries of the diagonal matrix are given by:
\begin{align}
    d_i = \sum_j a_{ij} = \mu(N-1) + \epsilon_i,
\end{align}
where $\epsilon_i \sim \mathcal{N} (0, \sigma_{diag}^2)$, and $\sigma_{diag} = \sigma \sqrt{N - 1}$. Note that the standard deviation $\sigma_{diag}$ of elements $d_i$ on the diagonal is by a factor of $\sqrt{N - 1}$ larger than that of the non-diagonal elements $a_{ij}$. However, for sufficiently large $N$, the deviation is significantly smaller than the sum of the expected magnitudes of the non-diagonal elements in that row:
\begin{align}
    \sigma_{diag} = \sigma \sqrt{N - 1} \ll& \sigma \sqrt{\frac{2}{\pi}} (N - 1) \\
    \leq& \sigma \sqrt{\frac{2}{\pi}} (N - 1) e^{-\mu^2 / 2\sigma^2} \\&\quad\quad+ \mu (N - 1)\left(1 - 2\Phi\left(-\frac{\mu}{\sigma}\right)\right) \nonumber \\
    =& \sum_{j\neq i} \mathrm{E}\left[|a_{ij}|\right].
\end{align}
Here, $\Phi$ denotes the normal cumulative distribution function. Consequently, the Laplacian $\mathbf{L}$ can be approximated by a matrix $\mathbf{L}'$, which is defined as follows:
\begin{align}
    \mathbf{L} &= \mathbf{D} - \mathbf{A}\\
    &= \mu (N - 1) \mathbbm{1} + \mathrm{diag}(\epsilon_1, ..., \epsilon_N) - \mathbf{A}\\
    &\approx \mu (N - 1) \mathbbm{1} - \mathbf{A}
    \eqqcolon \mathbf{L}'.
\end{align}

Crucially, as eigenvectors $\mathbf{u_k}$ of $\mathbf{A}$ are also eigenvectors of $\mathbf{L}'$,
\begin{align}
    \mathbf{L}' \mathbf{u_k} &= \mu (N - 1) \mathbf{u_k} - \mathbf{A} \mathbf{u_k} = \lambda_k' \mathbf{u_k},
\end{align}
the GFT based on the adjacency matrix and the Laplacian matrix will be similar.

The normality assumption can be a good approximation for highly interconnected networks with negative weights. The Pearson correlation graphs retrieved from the simulated data in the experimental section of this book chapter fulfils both conditions. For disconnected networks, the assumption is not met as weights between disconnected parts of the graph are zero and thereby not randomly sampled. 
Considering brain networks, certain pathological conditions, such as split-brain, can cause the brain graph to be disconnected. 
On the other hand, graphs with only positive graph weights can equally violate the normality assumption if the mean is not much larger than the standard deviation, as this gives rise to a skewed distribution. Many graph retrieval methods yield graphs with only positive weights.
The topic of negative graph weights is further explored in subsection \ref{ssec:negative}.

\subsection{Link to principal component analysis (PCA)}
\label{ssec:link_pca}
PCA projects multivariate samples onto a set of orthogonal components, which are the eigenvectors of the covariance matrix of the data set.
Crucially, if the covariance matrix is used for the GFT, then GFT and PCA are mathematically equivalent.
Note that for normalised signals with standard deviation one, which is common for some neurophysiological signals such as EEG signals, the correlation and the covariance matrix are the same. 
Note also that the diagonal entries of the signal correlation matrix are one and will not affect the 
\begin{algorithm}[H]
\caption{Dynamic neuroimaging data generation}
\label{alg:sim_EEG}
\hspace*{\algorithmicindent} \textbf{Input} $N_t$,\,$\mathbf{A}_s$,\,$h$,\,$\alpha$,\,$\beta$,\,$\gamma$\\
\hspace*{\algorithmicindent} \textbf{Output} $\mathbf{X}$
\begin{algorithmic}[1]
\State $\epsilon_{ij}, \tilde{\epsilon}_{ij} \sim \mathcal{N}(0, 1^2)$,\quad$i=1,...,N_c$,\:$j=1,...,N_t$
\State $\bm{\widehat{\epsilon}}_{i*} \gets \left(\mathcal{F}^{-1} \circ h \circ \mathcal{F}\right)(\bm{\tilde{\epsilon}}_{i*})$,\quad$i=1,...,N_c$
\State $\mathbf{x}_{*1}\gets \beta\bm{\widehat{\epsilon}}_{*1}$
\For{$t\gets 2, N_t$}
\State $\mathbf{\widehat{x}}_{*t-1} \gets \mathbf{x}_{*t-1} - \bar{\mathbf{x}}_{*t-1} +  \gamma \bm{\epsilon}_{*t}$ \label{algoline:timestep1}
\State $\mathbf{x}_{*t}\gets \alpha \mathbf{A}_s \mathbf{\widehat{x}}_{*t-1} + \beta\bm{\widehat{\epsilon}}_{*t}$ \label{algoline:timestep2}
\EndFor
\end{algorithmic}
\end{algorithm}
eigenvectors.
In other words, using the correlation matrix of normalised signals for GFT, either with or without diagonal elements, is equivalent to PCA.

\subsection{Negative graph weights}
\label{ssec:negative}

Some graph retrieval methods, such as the raw pairwise Pearson correlation, can generate negative weights in the adjacency matrix. 
This section will discuss some of the mathematical implications of negative weights for GSP.

A first consequence of using a graph with negative weights is that the Laplacian matrix is not necessarily diagonally dominant, which in turn means that the Laplacian can have negative eigenvalues. A downside of this is that the constant, or DC, graph Fourier mode is no longer the lowest Fourier mode, i.d., the mode with the lowest frequency.
An example of this can be seen in Figure \ref{fig:graph_Fourier_modes}, where the graph Fourier modes for two graphs are given, namely the geometric graph with only positive weights and the Pearson correlation graph with both positive and negative weights. The DC graph Fourier mode is the lowest mode for the geometric graph, which is not the case for the Pearson correlation graph.
The notion of frequency is related to the total variation, as defined in subsection \ref{ssec:total_variation}. Allowing negative weights means that the edge-based total variation, as defined in equation (\ref{eq:TV_edge}), can become negative. This may be 
A third consequence of negative weights is that they can cause negative entries in the degree matrix $\mathbf{D}$, which means that the square root of the inverse of $\mathbf{D}$, $\mathbf{D}^{-1/2}$, is not real-valued. As a result of this, the symmetric normalised Laplacian $\mathbf{L}_\mathrm{norm}=\mathbf{D}^{-1/2}\mathbf{L}\mathbf{D}^{-1/2}$, which may be required for certain applications, can not be computed.
While negative weights may be less intuitive, they can extend the applicability of graphs \cite{sedgewick2003algorithms}. In neuroimaging, for example, negative weights can be used to model inhibitory pathways in the brain \cite{wang2022networks}.

\section{Methodology}
\label{sec:methodology}
\subsection{Simulated data set}
\label{ssec:simulation}

We developed Algorithm \ref{alg:sim_EEG} to generate multivariate signal samples $\mathbf{X}_{sim}\in \mathbb{R}^{N_c\times N_t}$ with $N_c$ channels and $N_t$ time samples.
The generated samples simulate neuroimaging measurements, such as EEG, with which they share two crucial characteristics: Firstly, the samples have an underlying graph structure, such that each pair of signals has a specific connectivity associated to it. Secondly, the temporal signals have a specific spectral profile. 

The algorithm takes as input the number of generated time samples $N_t$, the weighted adjacency matrix $\mathbf{A}_s\in \mathbb{R}^{N_c\times N_c}$, the filter function $h$, as well as parameters $\alpha$, $\beta$ and $\gamma$.
While $\mathbf{A}_s$ controls the connectivity, or spatial structure, the filter function $h$ controls the spectral profile. Specifically, using two filter functions $h_1$ and $h_2$ allows to generate data samples with two conditions. The similarity between the two filter functions controls the difficulty of classifying the conditions. This simulates, for example, EEG data where multiple conditions, such as Alzheimer's disease or healthy control, each have a specific spectral profile \cite{jeong2004eeg}. Parameter $\alpha$ controls the strength of the correlation structure, while parameter $\beta$ controls the strength of the spectral structure. Parameter $\gamma$ controls the self-amplification of the signals during the simulation.

\begin{figure*}[t]
    \fontfamily{\sfdefault}\selectfont
    \begin{subfigure}[c]{0.48\textwidth}
    \centering
    \begin{tikzpicture}
    \draw (0, 0) node[inner sep=0] {\includegraphics[trim={0 4.7cm 0 0},clip,width=0.9\textwidth]{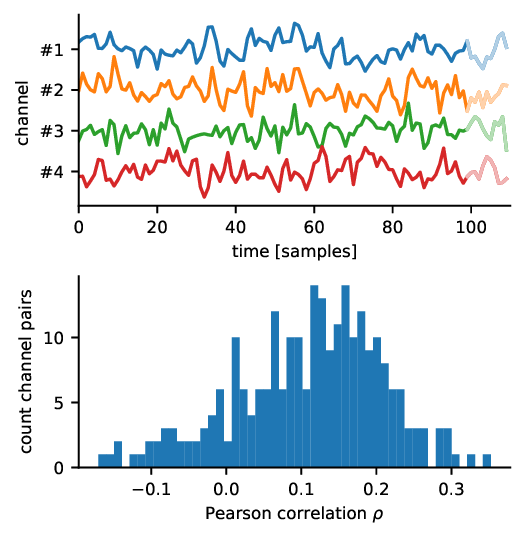}};
    \draw (-3.0, 1.8) node {A};
    \end{tikzpicture}
    \label{fig:time_signal}
    \end{subfigure}\hfill
    \begin{subfigure}[c]{0.48\textwidth}
    \centering
    \begin{tikzpicture}
    \draw (0, 0) node[inner sep=0] {\includegraphics[trim={0 5.1cm 0 0},clip,width=0.9\textwidth]{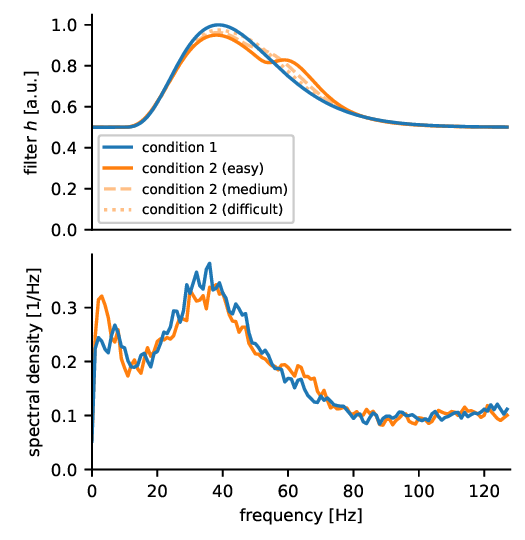}};
    \draw (-2.9, 1.8) node {C};
    \end{tikzpicture}
    \label{fig:WPSD_preset}
    \end{subfigure}
    \begin{subfigure}[c]{0.48\textwidth}
    \centering
    \begin{tikzpicture}
    \draw (0, 0) node[inner sep=0] {\includegraphics[trim={0 0 0 4.6cm},clip,width=0.9\textwidth]{figures/simulated_signals_and_channel_weights_distribution.eps}};
    \draw (-3.0, 2.0) node {B};
    \end{tikzpicture}
    \label{fig:correlation_distribution}
    \end{subfigure}\hfill
    \begin{subfigure}[c]{0.48\textwidth}
    \centering
    \begin{tikzpicture}
    \draw (0, 0) node[inner sep=0] {\includegraphics[trim={0 0 0 4.1cm},clip,width=0.9\textwidth]{figures/control_spectral_density.eps}};
    \draw (-2.9, 2.0) node {D};
    \end{tikzpicture}
    \label{fig:WPSD_sim_data}
    \end{subfigure}
    \caption{Simulated neurophysiological signals. (A) Time signals for four selected channels across the first 100 samples. Channels \texttt{\#}2, \texttt{\#}3 and \texttt{\#}4 are positively, weakly and negatively correlated to channel \texttt{\#}1, respectively. (B) Distribution of correlations of channel pairs, exhibiting the spatial connectivity structure. In the simulation, this structure is controlled by the matrix $\mathbf{A_s}$, but it is also affected by the simulation parameters. (C,D) Demonstration of power spectral density control and adjustment of classification difficulty. (C) The Fourier filter function $h$ used to colour the noise in Algorithm \ref{alg:sim_EEG}, assuming a sampling frequency of 256\,Hz. For each data set, two conditions are simulated. Condition 2 can be varied to make it easy (solid), medium (dashed), or difficult (dotted) to distinguish from condition 1. The difficulty depends on the similarity between the two conditions. (D) Welch power spectral density of simulated signals averaged across all channels for two easily distinguishable conditions. Figures (C) and (D) clearly show that the shape of the power spectral density profile of the simulated signals can be controlled. Parameter $\alpha$ in Algorithm \ref{alg:sim_EEG} can be used to reduce the power density at lower frequencies}
\label{fig:signal_characterisation}
\end{figure*}

In each time step of Algorithm \ref{alg:sim_EEG}, we firstly centre the graph signal of the previous time step $\mathbf{x}_{*t-1}$ around zero and add Gaussian noise $\gamma \bm{\epsilon}_{*t}$ to this signal (line \ref{algoline:timestep1}). Secondly, we use the adjacency matrix $\mathbf{A}_s$ as a GSO to translate the graph signal in time, which enforces the structural connectivity $\mathbf{A}_s$ in our data (line \ref{algoline:timestep2}). Thirdly, we scale the translated signal by $\alpha$ and add normalised coloured noise $\bm{\widehat{\epsilon}}_{*t}$ scaled with $\beta$, whose spectral density profile is controlled by the filter $h$ (line \ref{algoline:timestep2}). Finally, the simulated multivariate signal is labelled with its condition.

We used three different filter functions $h_2$ for condition 2 (orange lines in Figure \ref{fig:signal_characterisation}(c)), resulting in overall three data sets which are either easy, medium or difficult to classify.
The connectivity structure matrix $\mathbf{A}_s \in \mathbb{R}^{23\times 23}$ was generated by drawing weights $a^s_{ij}\sim \mathcal{U}_{[-0.1, 0.4]}$ from a uniform distribution for $i>j$, and setting $a^s_{ii} = 0.4$ and $a^s_{ji} = a^s_{ij}$. We set the simulation parameters to $\alpha = 0.5$, $\beta = 1$, and $\gamma = 1$.
For each simulated participant, we generated $N_c=23$ time-series signals with $N_t=2048$ time samples each.

Figure \ref{fig:signal_characterisation} shows the simulated signals (a) along with their spatial (b) and spectral (d) structure.
The spectral structure (d) of the simulated signals is enforced by the spectral density profile of the coloured noise (c), demonstrating that the spectral structure can be controlled.

\subsection{Analysis}
\label{ssec:analysis}
The goal of the analysis is to GFT-transform the multivariate signal into graph frequency signals and subsequently classify the samples into the two conditions, using only one graph frequency signal at a time. The transformed graph frequency signals may have features either not or only weakly present in the time signals. On the other hand, they may capture spectral features in multiple time signals, thereby allowing to reduce the dimensionality of the problem by limiting the number of analysed signals.
The structure of our graph frequency analysis is illustrated in Figure \ref{fig:analysis}. The approach allows us to determine the classification performance for each graph frequency, which we use to assess the quality of the dynamical structures in this graph frequency signal. 
\begin{figure*}[tbh]
    \centering
    \resizebox{\textwidth}{!}{%
    \input{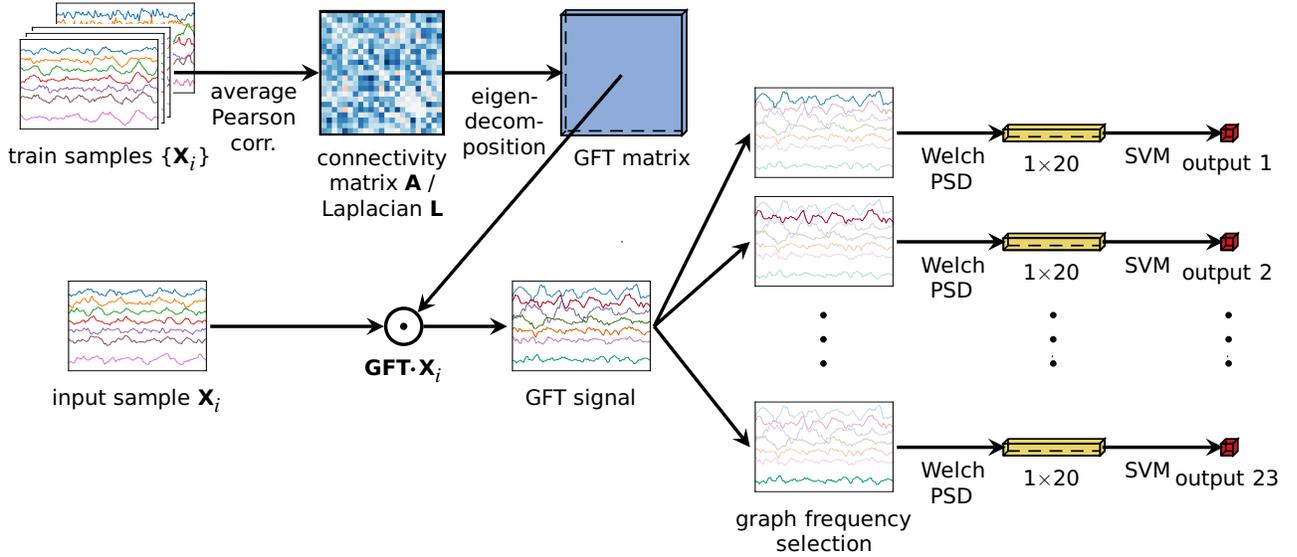}%
    }
    \setcounter{figure}{4}
    \vspace{-0.5cm}
    \caption{Illustration of the graph frequency analysis. For each cross-validation iteration, all simulated samples from the training set are used to construct either the connectivity matrix $\mathbf{A}$ or the Laplacian matrix $\mathbf{L}$, from which the GFT matrix is computed as the eigendecomposition (see subsection \ref{ssec:GFT}). Using the GFT matrix, the input sample $\mathbf{X}_i$ is transformed to its GFT signal $\tilde{\mathbf{X}}_i$. The signal is further split into the 23 graph frequency signals. Lastly, a support vector machine classifier is trained on the 20 time spectral features extracted from each graph frequency signal using Welch's power spectral density method. The performance of each graph frequency can then be used to assess the quality of the spectral features in the graph frequency signal}
    \label{fig:analysis}
\end{figure*}

The first step of the analysis consists of retrieving the graph structure from the training samples, for which we used the functional connectivity. Note that we need to use the same graph for all samples to keep the graph Fourier modes constant. Specifically, we computed the correlation matrix for each sample in the training set and subsequently averaged all matrices, yielding a common weighted adjacency matrix. 
Secondly, we carried out the GFT, yielding $N_c=23$ graph frequency signals. We used the weighted adjacency and the Laplacian matrix for the GFT in two separate experiments.
Thirdly, we extracted spectral features from those signals for each sample. To this end, we computed the Welch power spectral density with a window of 128 for each transformed signal, removed the last 28 values, and downsampled the remaining 100 values by averaging five values each, yielding 20 features per graph frequency.
Lastly, we trained a support vector machine classifier separately for each graph frequency to classify the labelled samples, using only the 20 features calculated from the graph frequency signal.
The baseline models are analysed following the same steps, except that the graph in the GFT was replaced as described in subsection \ref{ssec:baseline_models}.

\subsection{Testing}
\label{ssec:testing}
\begin{figure}[tb]
    \centering
    \resizebox{\columnwidth}{!}{


\renewcommand*\familydefault{\sfdefault} 
\fontfamily{\sfdefault}\selectfont

\definecolor{darkblue}{HTML}{1f4e79}
\definecolor{lightblue}{HTML}{00b0f0}
\definecolor{salmon}{HTML}{ff9c6b}
\definecolor{dodgerblue}{rgb}{0.12, 0.56, 1.0}
\definecolor{frenchblue}{rgb}{0.0, 0.45, 0.73}
\definecolor{green(pigment)}{rgb}{0.0, 0.65, 0.31}
\definecolor{macaroniandcheese}{rgb}{1.0, 0.74, 0.53}
\definecolor{glaucous}{rgb}{0.38, 0.51, 0.71}
\definecolor{hanblue}{rgb}{0.27, 0.42, 0.81}
\definecolor{newblue}{rgb}{0.56, 0.67, 0.85}
\definecolor{newgreen}{rgb}{0.67, 0.82, 0.57}
\definecolor{fireenginered}{rgb}{0.81, 0.09, 0.13}

\makeatletter
\pgfkeys{/pgf/.cd,
  parallelepiped offset x/.initial=2mm,
  parallelepiped offset y/.initial=2mm
}
\pgfdeclareshape{parallelepiped}
{
  \inheritsavedanchors[from=rectangle] 
  \inheritanchorborder[from=rectangle]
  \inheritanchor[from=rectangle]{north}
  \inheritanchor[from=rectangle]{north west}
  \inheritanchor[from=rectangle]{north east}
  \inheritanchor[from=rectangle]{center}
  \inheritanchor[from=rectangle]{west}
  \inheritanchor[from=rectangle]{east}
  \inheritanchor[from=rectangle]{mid}
  \inheritanchor[from=rectangle]{mid west}
  \inheritanchor[from=rectangle]{mid east}
  \inheritanchor[from=rectangle]{base}
  \inheritanchor[from=rectangle]{base west}
  \inheritanchor[from=rectangle]{base east}
  \inheritanchor[from=rectangle]{south}
  \inheritanchor[from=rectangle]{south west}
  \inheritanchor[from=rectangle]{south east}
  \backgroundpath{
    \southwest \pgf@xa=\pgf@x \pgf@ya=\pgf@y
    \northeast \pgf@xb=\pgf@x \pgf@yb=\pgf@y
    \pgfmathsetlength\pgfutil@tempdima{\pgfkeysvalueof{/pgf/parallelepiped
      offset x}}
    \pgfmathsetlength\pgfutil@tempdimb{\pgfkeysvalueof{/pgf/parallelepiped
      offset y}}
    \def\ppd@offset{\pgfpoint{\pgfutil@tempdima}{\pgfutil@tempdimb}}
    \pgfpathmoveto{\pgfqpoint{\pgf@xa}{\pgf@ya}}
    \pgfpathlineto{\pgfqpoint{\pgf@xb}{\pgf@ya}}
    \pgfpathlineto{\pgfqpoint{\pgf@xb}{\pgf@yb}}
    \pgfpathlineto{\pgfqpoint{\pgf@xa}{\pgf@yb}}
    \pgfpathclose
    \pgfpathmoveto{\pgfqpoint{\pgf@xb}{\pgf@ya}}
    \pgfpathlineto{\pgfpointadd{\pgfpoint{\pgf@xb}{\pgf@ya}}{\ppd@offset}}
    \pgfpathlineto{\pgfpointadd{\pgfpoint{\pgf@xb}{\pgf@yb}}{\ppd@offset}}
    \pgfpathlineto{\pgfpointadd{\pgfpoint{\pgf@xa}{\pgf@yb}}{\ppd@offset}}
    \pgfpathlineto{\pgfqpoint{\pgf@xa}{\pgf@yb}}
    \pgfpathmoveto{\pgfqpoint{\pgf@xb}{\pgf@yb}}
    \pgfpathlineto{\pgfpointadd{\pgfpoint{\pgf@xb}{\pgf@yb}}{\ppd@offset}}
  }
}
\makeatother

\tikzset{
  train/.style={
    parallelepiped,fill=white, draw,
    minimum width=1.6cm,
    minimum height=1.6cm,
    parallelepiped offset x=0.4cm,
    parallelepiped offset y=0.4cm,
    path picture={
      \draw[top color=newblue,bottom color=newblue]
        (path picture bounding box.south west) rectangle 
        (path picture bounding box.north east);
    },
    text=black,
  },
  test/.style={
    parallelepiped,fill=white, draw,
    minimum width=1.6cm,
    minimum height=1.6cm,
    parallelepiped offset x=0.4cm,
    parallelepiped offset y=0.4cm,
    path picture={
      \draw[top color=newgreen,bottom color=newgreen]
        (path picture bounding box.south west) rectangle 
        (path picture bounding box.north east);
    },
    text=black,
  },
}
\hspace{-3.75cm}
\noindent
\begin{tikzpicture}
  \node[test](test_1_1){{\Large Test}};
  \node[test, right=0.cm of test_1_1, yshift=-2.3cm](test_2_1){{\Large Test}};
  \node[test, right=0.cm of test_2_1, yshift=-2.3cm](test_3_1){{\Large Test}};
  \node[train, right=0.cm of test_3_1](train_3_1){{}};
  \node[train, right=0.cm of train_3_1](train_3_2){{}};
  \node[train, right=0.cm of train_3_2](train_3_3){{}};
  \node[train, right=0.cm of train_3_3](train_3_4){{}};
  \node[train, right=0.cm of train_3_4](train_3_5){{\Large Train}};
  \node[train, right=0.cm of train_3_5](train_3_6){{}};
  \node[train, right=0.cm of train_3_6](train_3_7){{}};
  \node[train, right=0.cm of train_3_7](train_3_8){{}};
  \node[train, right=0.cm of train_3_8](train_3_9){{\Large {\color{fireenginered}New}}};
  \node[train, right=0.cm of test_2_1](train_2_1){{}};
  \node[train, right=0.cm of train_2_1](train_2_2){{}};
  \node[train, right=0.cm of train_2_2](train_2_3){{}};
  \node[train, right=0.cm of train_2_3](train_2_4){{}};
  \node[train, right=0.cm of train_2_4](train_2_5){{\Large Train}};
  \node[train, right=0.cm of train_2_5](train_2_6){{}};
  \node[train, right=0.cm of train_2_6](train_2_7){{}};
  \node[train, right=0.cm of train_2_7](train_2_8){{}};
  \node[train, right=0.cm of train_2_8](train_2_9){{\Large {\color{fireenginered}New}}};
  \node[train, right=0.cm of test_1_1](train_1_1){{}};
  \node[train, right=0.cm of train_1_1](train_1_2){{}};
  \node[train, right=0.cm of train_1_2](train_1_3){{}};
  \node[train, right=0.cm of train_1_3](train_1_4){{}};
  \node[train, right=0.cm of train_1_4](train_1_5){{\Large Train}};
  \node[train, right=0.cm of train_1_5](train_1_6){{}};
  \node[train, right=0.cm of train_1_6](train_1_7){{}};
  \node[train, right=0.cm of train_1_7](train_1_8){{}};
  \node[train, right=0.cm of train_1_8](train_1_9){{}};
  
  \node at (9.7cm,-6.3cm) [circle,fill,inner sep=1.5pt]{};
  \node at (9.7cm,-6.7cm) [circle,fill,inner sep=1.5pt]{};
  \node at (9.7cm,-7.1cm) [circle,fill,inner sep=1.5pt]{};
  \node at (-0.8cm,1.2cm){};
  \node at (9.7cm,-7.5cm){};
  \node at (19,0.cm){};
  
  \node at (-1.3cm,0.cm)[align=left, text width=3.5cm]{\LARGE 1. iter.};
  \node at (-1.3cm,-2.3cm)[align=left, text width=3.5cm]{\LARGE 2. iteration};
  \node at (-1.3cm,-4.6cm)[align=left, text width=3.5cm]{\LARGE 3. iteration};

\end{tikzpicture}

}
    \caption{Illustration of the perpetual cross-validation used during the analysis. The whole data set is split into $k=10$ folds. One fold is assigned to be the testing set, while the remaining $k-1$ folds comprise the training set. For the subsequent iteration, a fold with newly generated data is added to the training set, while the testing fold is shifted by one fold}
    \label{fig:cross_validation}
\end{figure}
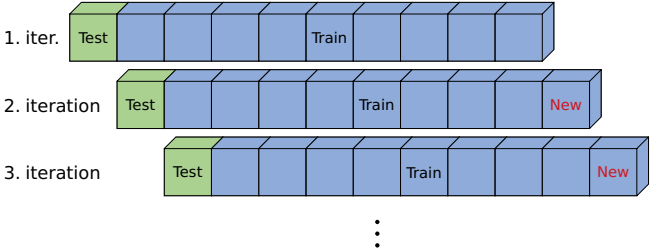

Generating the data matrices is time-consuming, as one simulation step is needed for each time sample. We therefore used a modified, perpetual version of cross-validation, illustrated in Figure \ref{fig:cross_validation}. Initially, $N_s=100$ samples are generated, divided evenly in condition 1 and condition 2. This data set is split into $k=10$ folds, nine of which are used for training. This results in 90 training samples, mimicking the sparsity of samples in neuroimaging. The remaining fold is used for testing, resulting in 10 testing samples per iteration. For the subsequent iteration, data is generated to form a new fold with $N_s / k = 10$ samples, which is added to the training set, whereas the testing fold is shifted by one fold. The final accuracy scores are averaged across all samples in the testing set and across all iterations. As a result of varying the number of iterations per model, our main model was tested on overall 5000 testing samples, whereas the permuted nodes GSP, random graph GSP, and single channel baseline model were tested on 5000, 4000, and 2000 testing samples, respectively.

The advantage of this perpetual version of cross-validation is that it can have arbitrarily many iterations, while at the same time reusing each fold $k$-times. It further does not require much storage, as used samples can be overwritten with newly generated samples.

\subsection{Baseline models}
\label{ssec:baseline_models}
We compose a set of three baseline models, which either distort the graph in GSP, or use conventional signal processing. The goal of all three models is to be able to attribute the boost in performance to using the connectivity structure underlying the data.

The first baseline model is the \textit{permuted nodes GSP model}, which uses the graph underlying the data, but randomly permutes the nodes of this graph.
In this way, the eigenmodes have the same weights as the actual model, but at different locations. Therefore, the model distorts the graph structure, while retaining the effect of the weight magnitudes.
The second, related baseline model is the \textit{random graph GSP model}, which uses GSP with a randomly generated graph. While this model transforms the data using GFT, these transformations are not based on the actual graph structure.
The third baseline model is the \textit{single channel model}, which does not transform the data and is equivalent to conventional signal processing. It can also be viewed as GSP with the identity transform, which is given by the identity matrix $\mathbbm{1}$.
Note that the "graph Fourier modes" are given by the single channels and have no natural ordering, which can also be seen by the fact that the eigenvalues of the identity matrix are degenerate. This model is the only model that excludes the graph structure, making it an indispensable baseline model.

\section{Results}
\label{sec:results}

\begin{figure}[tb!]
    \centering
    \resizebox{\columnwidth}{!}{\includegraphics[trim={0.25cm 0 0 0},clip]{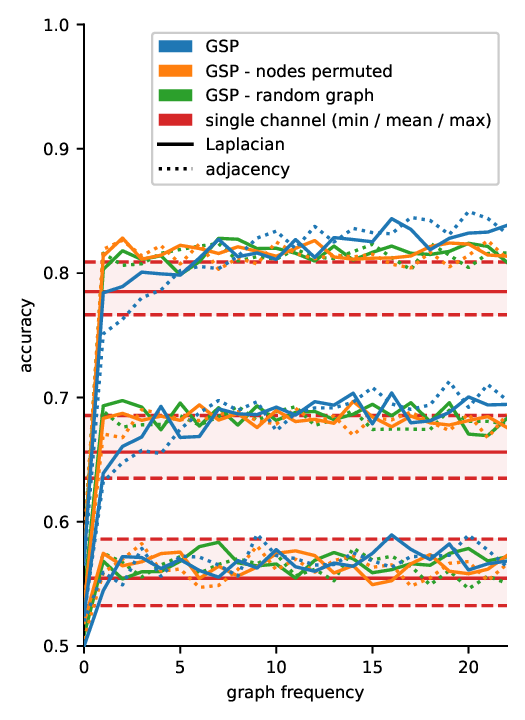}}
    \caption{Classification accuracy as a function of the graph frequency of the transformed signals. Three data sets with varying classification difficulty were simulated, whereby the difficulty was controlled by modifying the filter $h$ as shown in Figure \ref{fig:signal_characterisation}(c). Easy, medium, and difficult classification difficulties result in the high, medium, and low accuracies overlaid in the figure, respectively. The model is shown in blue, while the baseline models are shown in orange, green and red. Note that the single channel baseline model in red does not have a graph frequency ordering. Models using the GFT with the Laplacian (adjacency) matrix are shown in solid (dotted) lines}
    \label{fig:acc_vs_eigen}
\end{figure}

Figure \ref{fig:acc_vs_eigen} illustrates the results of our analysis of the artificially generated data. Specifically, it shows the classification accuracy in dependence on the graph frequency of the transformed signal for three simulated difficulty levels. The main model, shown in blue, is compared against the three baseline models. The accuracy scores in the figures are averaged across all testing samples, as described in subsection \ref{ssec:testing}.

The accuracy of the graph DC component is by far the lowest. This can be understood by considering the case of the Laplacian matrix: When using this matrix for the GFT, the graph DC component is equivalent to the average across all signals. However, when averaging the signals, some temporal structures in the signals are also averaged out.

The model performs worse at lower graph frequencies than all GSP-based baseline models, and only equally well than the single channel baseline model.
The relatively poor performance of the single channel baseline model can be attributed to the fact that, as opposed to all other models, this model does not gather spectral structures from more than one signal, giving the classifier less information to use.
For easy to classify data sets, our model perform slightly better than the baseline models at higher graph frequencies.

The choice of the specific graph representation for the GFT does not have a strong impact on the results, which has been reported earlier \cite{huang2018graph_perspective}.
Further, we show that our results are replicated for various difficulty levels.

\section{Discussion and Conclusion}
\label{sec:discussion_conclusion}
In the experimental section of this book chapter, we have developed a neuroimaging data generation algorithm, which dynamically generates arbitrarily many samples.
We have demonstrated that this data has a spatial connectivity structure, as well as a controllable spectral structure.
However, our simulated data may differ from real data in crucial ways.
For example, the spectral structure is enforced by adding coloured noise, and does not primarily arise out of the network interaction, which however may be the case for real-life neuroimaging data. 
Nevertheless, we uphold that the generated data capture the essential characterisics of neurophysiological signals relevant for GSP, as demonstrated in Figure \ref{fig:signal_characterisation}, and hence that their analysis allows to draw some general conclusions about GSP applied to real-life neurophysiological signals.

One limitation of GSP for neuroimaging is the ambiguity of the GFT, which is due to the number of choices for how to retrieve the graph (see subsection \ref{ssec:graph_retrieval}) and for which graph representation to use.
These limitations, in combination with the novelty of the method, highlight the need for a thorough validation procedure for GSP, as opposed to basing it mainly on theoretical justifications. To this end, we have introduced a baseline testing framework for GSP in subsection \ref{ssec:baseline_models}, consisting of a set of overall three baseline models. Note that the GFT can be linked to the PCA, as shown in subsection \ref{ssec:link_pca}, making the PCA unsuitable as a baseline model.

Using the artificial data, we have systematically evaluated the classification performance of single graph frequency signals in terms of their graph frequency. 
We found that higher graph frequencies outperform lower ones, which is in line with the results obtained in \cite{menoret2017evaluating}. This specific result may not generalise to all neurophysiological data sets, simulated or real-life, because it may depend on the precise interaction between the spatial and the spectral structure. The result is nevertheless surprising, as we expected lower graph frequencies to preserve temporal structures better, given that they are composed of more closely related channels. 
We further found that even at higher graph frequencies our model performs only marginally better than the baseline models. Note also that higher graph frequencies may not relate much to the spatial structure, as pointed out in subsection \ref{ssec:GFT}. 
Taken together, our results do not suggest that GSP leverages the spatial structure to improve spectral feature-based classification.

One possible explanation for this null result is that the GFT linearly mixes the time signals, which each have different phases \cite{thatcher2012coherence}. The frequencies in the time signals can interfere due to the phase differences, leading to an attenuation of the spectral features. Other GSP applications that rely on combining the time signals linearly may be equally affected by this interference effect, such as graph filtering.
This shortcoming may limit the application of GSP for EEG, where the dynamic structure, i.e., the spectral features, is vital to characterise the data, but it may also impose limitations for applying GSP to neurophysiological data in general.

\section*{Acknowledgements}
The authors would like to thank Dominik Klepl for important discussions.

\newpage
\bibliographystyle{unsrt}
\bibliography{IEEESPMB}

\appendix
\renewcommand\thefigure{\thesection.\arabic{figure}} 
\renewcommand\thetable{\thesection.\Roman{table}}

\end{document}